# The spectrum of coupled random matrices

By M. ADLER and P. VAN MOERBEKE*

Table of Contents

## 0. Introduction

The study of the spectrum of coupled random matrices has received rather little attention. To the best of our knowledge, coupled random matrices have been studied, to some extent, by Mehta in [16], [17], [11]. In this work, we explain how the integrable technology can be brought to bear to gain insight into the nature of the distribution of the spectrum of coupled Hermitean random matrices and the equations the associated probabilities satisfy. In particular,

*The first author gratefully acknowledges the support of a National Science Foundation grant # DMS-9503246. The second author gratefully acknowledges the support of a National Science Foundation grant # DMS-9503246, a Nato, a FNRS and a Francqui Foundation grant.

the two-Toda lattice, its algebra of symmetries and its vertex operators will play a prominent role in this interaction. Namely, the method is to introduce time parameters, in an artificial way, and to *dress up* a certain matrix integral with a *vertex integral operator*, for which we find Virasoro-like differential equations; for a state of the art survey of these methods, see [20]. These methods lead to very simple nonlinear third-order partial differential equations for the joint statistics of the spectra of two coupled Gaussian random matrices.

*Bi-orthogonal polynomials and the two-Toda lattice.* Given a weight, depending on parameters $t = (t_1, t_2, \ldots)$ and $s = (s_1, s_2, \ldots) \in \mathbb{C}^\infty$ and a coupling constant $c$,

$$\rho(x,y)dx\,dy = dx\,dy\, e^{\sum_1^\infty t_i x^i + cxy - \sum_1^\infty s_i y^i},$$

consider the associated (monic) bi-orthogonal polynomials $p_i^{(1)}(x)$ and $p_i^{(2)}(y)$ of degree $i$ defined by

$$\langle p_k^{(1)}, p_\ell^{(2)} \rangle = h_k \delta_{k,\ell}, \text{ where } \langle f, g \rangle = \iint_{\mathbb{R}^2} dx\,dy\, f(x)g(y)\rho(x,y).$$

In [2], it was shown that the associated pair of semi-infinite vectors, depending on $(t,s)$,

$$\Psi_1(z) := e^{\sum_1^\infty t_k z^k} p^{(1)}(z) \text{ and } \Psi_2^*(z) := e^{-\sum_1^\infty s_k z^{-k}} h^{-1} p^{(2)}(z^{-1}),$$

has the following expression[1]

$$(0.1) \quad \Psi_1(z) = \left( \frac{\tau_n(t - [z^{-1}], s)}{\tau_n(t,s)} e^{\sum_1^\infty t_i z^i} z^n \right)_{n \in \mathbb{Z}},$$

$$\Psi_2^*(z) = \left( \frac{\tau_n(t, s + [z])}{\tau_{n+1}(t,s)} e^{-\sum_1^\infty s_i z^{-i}} z^{-n} \right)_{n \in \mathbb{Z}},$$

in terms of the $2n$-uple integrals

$$(0.2) \quad \tau_n(t,s) := \iint_{\mathbb{R}^{2n}} d\vec{x}\, d\vec{y}\, \Delta(\vec{x}) \Delta(\vec{y}) \prod_{k=1}^n e^{\sum_1^\infty (t_i x_k^i - s_i y_k^i) + cx_k y_k},$$

which form a vector $\tau := (\tau_n)_{n \geq 0}$ of $\tau$-functions. This provides a concrete realization of the Sato representation of the two-Toda wave functions in terms of $\tau$-functions $\tau_n$; see [18], [2].

The pair of semi-infinite matrices $L := (L_1, L_2)$,[2] defined by:

$$(0.3) \quad z\Psi_1 = L_1 \Psi_1, \quad z^{-1}\Psi_2^* = L_2^\top \Psi_2^*,$$

---

[1] $[\alpha] = (\alpha, \alpha^2/2, \alpha^3/3, \ldots)$

[2] $L_1$ is lower-triangular, except for a subdiagonal just above the diagonal with all entries $= 1$ and $L_2$ is upper-triangular, except for a nonzero subdiagonal just below the diagonal.

together with the pair $(\Psi_1, \Psi_2^*)$, satisfy the two-Toda lattice equation; in particular,

$$(0.4) \qquad \frac{\partial L_i}{\partial t_n} = [(L_1^n)_u, L_i] \quad \text{and} \quad \frac{\partial L_i}{\partial s_n} = [(L_2^n)_\ell, L_i], \quad i = 1, 2,$$

for the usual Lie algebra splitting $(\ )_u, (\ )_\ell$, explained in Section 2. The familiar standard Toda lattice (on tridiagonal matrices) is a special reduction of the two-Toda lattice, in the same way that the Korteweg-de Vries equation (KdV) is a reduction of the Kadomtsev-Petviashvili equation (KP).

Conversely, starting from the the two-Toda lattice equations (0.4), one is led to wave functions $\Psi_1$ and $\Psi_2^*$ and a representation in terms of $\tau$-functions as in (0.1). As will be established in Section 3, the functions $\tau_n$ satisfy the standard KP-equation in $t$ and $s$ separately (see the beginning of §3), but they also satisfy another (new and useful) equation, which is third-order, relating $t$- and $s$-derivatives, namely:

THEOREM 0.1. *Two-Toda $\tau$-functions $\tau_n(t,s)$ satisfy:*[3]

$$\left\{\frac{\partial^2 \log \tau_n}{\partial t_1 \partial s_2}, \frac{\partial^2 \log \tau_n}{\partial t_1 \partial s_1}\right\}_{t_1} + \left\{\frac{\partial^2 \log \tau_n}{\partial s_1 \partial t_2}, \frac{\partial^2 \log \tau_n}{\partial t_1 \partial s_1}\right\}_{s_1} = 0.$$

*Vertex operators and "Christoffel-Darboux" kernels.* Bäcklund-Darboux transformations refer to the general recipe of factorizing differential or difference operators and flipping the factors, to form a new operator. When we let this situation flow in time, the new wave functions (eigenfunctions) can be expressed in terms of the old ones as Wronskians (continuous or discrete), and the new $\tau$-function is expressed in terms of the old ones, by means of vertex operators. So, the latter can be viewed as generators of Bäcklund-Darboux transformations for differential or difference operators at the level of $\tau$-functions. Typically, vertex operators $\mathbb{X}$ map $\tau$-functions into $\tau$-functions, and their squares vanish; although $\tau$ satisfies a highly nonlinear equation, vertex operators have an additive property; that is $\tau + \mathbb{X}\tau$ is a $\tau$-function as well!

With the two-Toda lattice, we associate *four* different vertex operators $\mathbb{X}_{ij}(\lambda, \mu)$, for $1 \leq i, j \leq 2$; they map infinite vectors of $\tau$-functions into $\tau$-vectors, as explained in Section 5. The vertex operators $\mathbb{X}_{11}$ and $\mathbb{X}_{22}$ are basic vertex operators for Toda, and KP, as well, whereas $\mathbb{X}_{12}$ and $\mathbb{X}_{21}$ are vertex operators, native to two-Toda. In particular, we construct

$$\mathbb{X}_{12}(\mu, \lambda) = \Lambda^{-1} \chi(\lambda) X(-s, \lambda) X(t, \mu) \chi(\mu),$$

---

[3]in terms of the Wronskian $\{f, g\}_t = \frac{\partial f}{\partial t} g - f \frac{\partial g}{\partial t}$.

with $\Lambda$ the customary shift-operator $(\Lambda v)_n = v_{n+1}$, and with

$$X(t, \lambda) := e^{\sum_1^\infty t_i \lambda^i} e^{-\sum_1^\infty \lambda^{-i} \frac{1}{i} \frac{\partial}{\partial t_i}}, \quad \chi(\lambda) := \mathrm{diag}(..., \lambda^{-1}, 1, \lambda, ...).$$

Besides this work, the vertex operator $\mathbb{X}_{12}$ will also play a major role in our later work on symmetric and symplectic matrix integrals. Given a two-Toda lattice $\tau$-vector $\tau = (...\tau_{-1}, \tau_0, \tau_1, ...)$, we have that $\tau + \mathbb{X}_{12}(y, z)\tau$ is another $\tau$-vector. But more is true. We show that the kernels $K_{12,n}(y, z)$, defined by the ratios $(\mathbb{X}_{12}\tau)_n/\tau_n$, have *eigenfunction expansions* in terms of the eigenfunctions $\Psi$, reminiscent of the Christoffel-Darboux formula for orthogonal polynomials; to be precise,

THEOREM 0.2. *We have (§5):*

$$(0.5) \qquad K_{12,n}(y, z) := \frac{1}{\tau_n} \mathbb{X}_{12}(y, z)\tau_n = \sum_{-\infty < j < n} \Psi_{2j}^*(z^{-1}) \Psi_{1j}(y),$$

*together with a Fredholm determinant-like formula,*

$$(0.6) \qquad \det\left((K_{ij,n}(y_\alpha, z_\beta))_{1 \le \alpha, \beta \le k}\right) = \frac{1}{\tau_n} \left(\prod_{\ell=1}^k \mathbb{X}_{ij}(y_\ell, z_\ell)\tau\right)_n.$$

*In the semi-infinite case, the sum in (0.5) is replaced by $\sum_{0 \le j < n}$.*

*Vertex operators, Virasoro algebras and two-Toda symmetries.* The vertex operators provide central extension realizations for Virasoro algebras; e.g.,
(0.7)
$$\frac{\partial}{\partial y} y^{k+1} \mathbb{X}_{12}(y, z) = [\mathbb{J}_k^{(2)}, \mathbb{X}_{12}(y, z)] \text{ and } \frac{\partial}{\partial z} z^{k+1} \mathbb{X}_{12}(y, z) = [\tilde{\mathbb{J}}_k^{(2)}, \mathbb{X}_{12}(y, z)]$$

with

$$\begin{aligned}(0.8) \qquad \mathbb{J}_k^{(2)} &:= \left(J_{k,n}^{(2)}\right)_{n \in \mathbb{Z}} \\ &= \frac{1}{2}\left(J_k^{(2)} + (2n + k + 1)J_k^{(1)} + n(n+1)J_k^{(0)}\right)_{n \in \mathbb{Z}} \\ \tilde{\mathbb{J}}_k^{(2)} &:= \left(\tilde{J}_{k,n}^{(2)}\right)_{n \in \mathbb{Z}} = \mathbb{J}_k^{(2)}\Big|_{t \mapsto -s}.\end{aligned}$$

forming Virasoro algebras of *central charge* $c = -2$.

In (0.8), $J_k^{(\ell)}$ equals $\delta_{k0}$ for $\ell = 0$, Heisenberg generators for $\ell = 1$ and Virasoro generators for $\ell = 2$. It follows that the vertex operators $\mathbb{X}_{ij}$ of the two-Toda lattice form, upon expansion, the generators of a large algebra of symmetries, which come from the master symmetries for the pair of matrices $L = (L_1, L_2)$.

This is a special case ($\alpha = 1$) of a more general statement concerning vector-vertex operators, depending on a parameter $\alpha$,

$$\mathbb{X}_\alpha(t, u) = \Lambda^{-1} e^{\sum_1^\infty t_i u^i} e^{-\alpha \sum_1^\infty \frac{u^{-i}}{i} \frac{\partial}{\partial t_i}} \chi(u^\alpha).$$

THEOREM 0.3. *The generators $\mathbb{J}_k^{(2)}(\alpha)$, defined by*

$$\frac{\partial}{\partial z} z^{k+1} \mathbb{X}_\alpha(t,z) = \left[\mathbb{J}_k^{(2)}(\alpha), \mathbb{X}_\alpha(t,z)\right],$$

*form a Virasoro algebra*

$$\left[\mathbb{J}_k^{(2)}(\alpha), \mathbb{J}_\ell^{(2)}(\alpha)\right] = (k-\ell)\mathbb{J}_{k+\ell}^{(2)}(\alpha) + c\left(\frac{k^3-k}{12}\right)\delta_{k,-\ell},$$

*with central charge*

$$c = 1 - 6\left(\left(\frac{\alpha}{2}\right)^{1/2} - \left(\frac{\alpha}{2}\right)^{-1/2}\right)^2.$$

*Commutation of Virasoro and "vertex integral operators".* Consider now a more general weight $\rho(y,z)dydz := \rho_{t,s}(y,z)dydz := e^{V_{t,s}(y,z)}dydz$ on $\mathbb{R}^2$, with $\rho_0 = e^{V_0}$, where

(0.9) $$\begin{aligned} V_{t,s}(y,z) &:= V_0(y,z) + \sum_1^\infty t_i y^i - \sum_1^\infty s_i z^i \\ &= \sum_{i,j\geq 1} c_{ij} y^i z^j + \sum_{i\geq 1} t_i y^i - \sum_{i\geq 1} s_i z^i, \end{aligned}$$

and a set $E \subset \mathbb{R}^2$ of the form,

(0.10) $$E = E_1 \times E_2 := \cup_{i=1}^r [a_{2i-1}, a_{2i}] \times \cup_{i=1}^s [b_{2i-1}, b_{2i}] \subset \mathbb{R}^2,$$

involving disjoint unions. The weight (0.9) and the boundary of the set (0.10) enable one to define two types of operators:

(i) Virasoro-like operators, for $k \geq -1$,

(0.11) $$\begin{aligned} \mathbb{V}_k : &= -\sum_{i=1}^r a_i^{k+1} \frac{\partial}{\partial a_i} + \mathbb{J}_k^{(2)} + \sum_{i,j\geq 1} ic_{ij} \frac{\partial}{\partial c_{i+k,j}} \\ \tilde{\mathbb{V}}_k : &= -\sum_{i=1}^r b_i^{k+1} \frac{\partial}{\partial b_i} + \tilde{\mathbb{J}}_k^{(2)} + \sum_{i,j\geq 1} jc_{ij} \frac{\partial}{\partial c_{i,j+k}}, \end{aligned}$$

with $\mathbb{J}_k^{(2)}$ and $\tilde{\mathbb{J}}_k^{(2)}$ as in (0.8).

(ii) An operator defined by a vertex integral operator $\mathbb{X}_{12}$ over $E$,

(0.12) $$\mathbb{U}_E := \iint_E dx\, dy\, \rho(x,y) \mathbb{X}_{12}(x,y).$$

It is an important ingredient in this work that $\mathbb{V}_k$ and $(\mathbb{U}_E)^n$ commute:

$$[\mathbb{V}_k, (\mathbb{U}_E)^n] = [\tilde{\mathbb{V}}_k, (\mathbb{U}_E)^n] = 0 \text{ for all } n \geq 1 \text{ and } k \geq -1.$$

*Two-matrix integrals over product sets and Virasoro constraints.* It follows that the $2n$-uple integral, like (0.2), but taken over the set $E^n = E_1^n \times E_2^n \subset \mathbb{R}^{2n}$,

$$(0.13) \quad \tau_n^E(t,s) := \int\int_{E^n} d\vec{x}\, d\vec{y}\, \Delta(\vec{x})\Delta(\vec{y}) \prod_{k=1}^n e^{\sum_1^\infty (t_i x_k^i - s_i y_k^i) + \sum_{i,j\geq 1} c_{ij} x_k^j y_k^j},$$

is related to $\tau := \left(\tau_n^{\mathbb{R}^2}\right)_{n\geq 0}$, by

$$\tau_n^E = ((\mathbb{U}_E)^n \tau)_n.$$

This implies, setting all $c_{ij} = 0$, but $c_{11} = c$:

THEOREM 0.4.  $\tau_n$ and $\tau_n^E$ satisfy the Virasoro-like partial differential equations, labeled by $k = -1, 0, 1, \ldots$:

$$(0.14) \quad \left(-\sum_{i=1}^r a_i^{k+1} \frac{\partial}{\partial a_i} + J_{k,n}^{(2)}\right) \tau_n^E + c\, p_{k+n}(\tilde{\partial}_t) p_n(-\tilde{\partial}_s) \tau_1^E \circ \tau_{n-1}^E = 0$$

$$\left(-\sum_{i=1}^s b_i^{k+1} \frac{\partial}{\partial b_i} + \tilde{J}_{k,n}^{(2)}\right) \tau_n^E + c\, p_n(\tilde{\partial}_t) p_{k+n}(-\tilde{\partial}_s) \tau_1^E \circ \tau_{n-1}^E = 0.$$

$J_{k,n}^{(2)}$ and $\tilde{J}_{k,n}^{(2)}$ were defined in (0.8), the Hirota symbols $p(\tilde{\partial}_t)q(-\tilde{\partial}_s)f \circ g$ in (6.1) and the $p_i$'s are the elementary Schur polynomials.

*Application to the spectrum of coupled random matrices.* Consider a product ensemble $(M_1, M_2) \in \mathcal{H}_n^2 := \mathcal{H}_n \times \mathcal{H}_n$ of $n \times n$ Hermitean matrices, equipped with a Gaussian probability measure,

$$(0.15) \quad c_n dM_1 dM_2\, e^{-\frac{1}{2}\mathrm{Tr}(M_1^2 + M_2^2 - 2cM_1 M_2)},$$

where $dM_1 dM_2$ is Haar measure on the product $\mathcal{H}_n^2$, with each $dM_i$,

$$(0.16) \quad dM_1 = \Delta_n^2(x) \prod_1^n dx_i\, dU \text{ and } dM_2 = \Delta_n^2(y) \prod_1^n dy_i\, dU$$

decomposed into radial and angular parts. We define differential operators $\mathcal{A}_k$, $\mathcal{B}_k$ of *weight $k$*, in terms of the coupling constant $c$, appearing in (0.15), and the boundary of the set

$$(0.17) \quad E = E_1 \times E_2 := \cup_{i=1}^r [a_{2i-1}, a_{2i}] \times \cup_{i=1}^s [b_{2i-1}, b_{2i}] \subset \mathbb{R}^2.$$

They form a closed Lie algebra, as spelled out in (11.4):

$$(0.18)$$
$$\mathcal{A}_1 = \frac{1}{c^2 - 1}\left(\sum_1^r \frac{\partial}{\partial a_j} + c\sum_1^s \frac{\partial}{\partial b_j}\right), \quad \mathcal{B}_1 = \frac{1}{1 - c^2}\left(c\sum_1^r \frac{\partial}{\partial a_j} + \sum_1^s \frac{\partial}{\partial b_j}\right),$$
$$\mathcal{A}_2 = \sum_{j=1}^r a_j \frac{\partial}{\partial a_j} - c\frac{\partial}{\partial c}, \qquad \mathcal{B}_2 = \sum_{j=1}^s b_j \frac{\partial}{\partial b_j} - c\frac{\partial}{\partial c}.$$

The following theorem deals with the joint distribution (§11),

(0.19) $\quad P_n(E) := P(\text{all}(M_1\text{-eigenvalues}) \in E_1, \ \text{all}(M_2\text{-eigenvalues}) \in E_2),$

and leads to a formula, which is the "mirror image" of Theorem 0.1.

THEOREM 0.5 (Gaussian probability). *The statistics (0.19) satisfies the $n$-independent nonlinear third-order partial differential equation*[4] *( $F_n := \frac{1}{n} \log P_n(E)$ ):*

(0.20)
$$\left\{\mathcal{B}_2\mathcal{A}_1 F_n, \ \mathcal{B}_1\mathcal{A}_1 F_n + \frac{c}{c^2-1}\right\}_{\mathcal{A}_1} - \left\{\mathcal{A}_2\mathcal{B}_1 F_n, \ \mathcal{A}_1\mathcal{B}_1 F_n + \frac{c}{c^2-1}\right\}_{\mathcal{B}_1} = 0.$$

*Remark* 1. Since the equation above for the joint statistics is independent of the size $n$, the same joint statistics for infinite coupled ensembles should presumably be given by the same partial differential equation.

*Remark* 2. For $E = E_1 \times E_2 := (-\infty, a] \times (-\infty, b]$, equation (0.20) takes on the following form: Upon introducing the new variables $x := -a + cb$, $y := -ac + b$, the differential operators $\mathcal{A}_1$ and $\mathcal{B}_1$ take on the simple form $\mathcal{A}_1 = \partial/\partial x$, $\mathcal{B}_1 = \partial/\partial y$ and (0.20) becomes

$$\frac{\partial}{\partial x}\left(\frac{(c^2-1)^2 \frac{\partial^2 F_n}{\partial x \partial c} + 2cx - (1+c^2)y}{(c^2-1)\frac{\partial^2 F_n}{\partial x \partial y} + c}\right) = \frac{\partial}{\partial y}\left(\frac{(c^2-1)^2 \frac{\partial^2 F_n}{\partial y \partial c} + 2cy - (1+c^2)x}{(c^2-1)\frac{\partial^2 F_n}{\partial y \partial x} + c}\right).$$

*Remark* 3. Equation (0.20) also has a "zero-curvature" formulation, namely:

(0.21) $\quad\quad\quad\quad [\mathcal{A}_1 - X_n, \mathcal{B}_1 - Y_n] = 0,$

with

$$X_n := \frac{\mathcal{A}_2\mathcal{B}_1 F_n}{\mathcal{A}_1\mathcal{B}_1 F_n + \frac{c}{c^2-1}} \quad \text{and} \quad Y_n := \frac{\mathcal{B}_2\mathcal{A}_1 F_n}{\mathcal{B}_1\mathcal{A}_1 F_n + \frac{c}{c^2-1}}.$$

The last section deals with coupled matrix ensembles, where the joint statistics is given by the "Laguerre distribution." Unlike Theorem 0.5, the Laguerre case for $n \times n$ matrices lead to (inductive) differential equations for the matrix integral (0.19); indeed, the equation contains a term, which is expressible in terms of the same expression for $(n-1) \times (n-1)$ matrices.

*Acknowledgment.* We thank Taka Shiota for many useful discussions concerning Fay identities. We also thank Edward Frenkel for urging us to compute the central charge for the Virasoro algebra defined in Theorem 0.3.

---

[4]in terms of the Wronskian $\{f,g\}_X = Xf.g - f.Xg$, with regard to a first order differential operator $X$.

## 1. Operators $\Lambda$ and $\varepsilon$ with $[\Lambda, \varepsilon] = 1$ and the $\delta$-function

Define the column vector $\chi(z) = (z^n)_{n \in \mathbb{Z}}$, and matrix operators $\Lambda$, $\Lambda^*$, $\varepsilon$, $\varepsilon^*$ as follows:

$$\Lambda \chi(z) = z\chi(z), \quad \varepsilon \chi(z) = \frac{\partial}{\partial z}\chi(z),$$

$$\Lambda^* \chi(z) = z^{-1}\chi(z), \quad \varepsilon^* \chi(z) = \frac{\partial}{\partial z^{-1}}\chi(z).$$

Note that

$$\Lambda^* = \Lambda^\top = \Lambda^{-1}, \quad \varepsilon^* = -\varepsilon^\top + \Lambda,$$

and

(1.1) $$\begin{cases} \Lambda^\top \chi(z^{-1}) = z\chi(z^{-1}), \ \Lambda\, \chi(z^{-1}) = z^{-1}\chi(z^{-1}) \\ \varepsilon^\top \chi(z^{-1}) = z^{-1}\chi(z^{-1}) - \frac{\partial}{\partial z}\chi(z^{-1}) \\ \varepsilon^{*\top} \chi(z^{-1}) = z\chi(z^{-1}) - \frac{\partial}{\partial z^{-1}}\chi(z^{-1}). \end{cases}$$

The operators $\Lambda, \Lambda^*, \varepsilon, \varepsilon^*$ have the following matrix representation:

(1.2) $$\begin{aligned} \Lambda &= (\delta_{i,j-1})_{i,j \in \mathbb{Z}}, & \varepsilon &= \operatorname{diag}(i) \cdot \Lambda^{-1} = (i\, \delta_{i,j+1})_{i,j \in \mathbb{Z}} \\ \Lambda^* &= (\delta_{i,j+1})_{i,j \in \mathbb{Z}}, & \varepsilon^* &= -\operatorname{diag}(i) \cdot \Lambda = (-i\delta_{i,j-1})_{i,j \in \mathbb{Z}}. \end{aligned}$$

For future use, we also introduce the $\delta$-function,

(1.3) $$\delta(t) = \sum_{n=-\infty}^{\infty} t^n = \frac{1}{1-t} + \frac{t^{-1}}{1-t^{-1}},$$

with the customary property

$$f(\lambda, \mu)\delta\left(\frac{\lambda}{\mu}\right) = f(\lambda, \lambda)\delta\left(\frac{\lambda}{\mu}\right).$$

Note the function

(1.4) $$\delta(\lambda, \mu) := \frac{1}{\mu}\delta\left(\frac{\lambda}{\mu}\right) = \frac{1}{\mu}\sum_{-\infty}^{\infty}\left(\frac{\lambda}{\mu}\right)^n$$

has the usual $\delta$-function property

$$\frac{1}{2\pi i}\oint f(\lambda, \mu)\delta(\lambda, \mu)d\mu = f(\lambda, \lambda)$$

and is a function of $\lambda - \mu$ only, since

(1.5) $$\left(\frac{\partial}{\partial \lambda} + \frac{\partial}{\partial \mu}\right)\delta(\lambda, \mu) = 0.$$

For future use, we state:

LEMMA 1.1 (T. Shiota). *We have the following matrix representation:*[5]

(1.6)
$$e^{(\mu-\lambda)\varepsilon}\delta(\lambda,\Lambda) = \frac{1}{\lambda}\chi(\mu)\otimes\chi^*(\lambda) \text{ and}$$
$$e^{(\mu-\lambda)\varepsilon^*}\delta(\lambda,\Lambda^*) = \frac{1}{\lambda}\chi^*(\mu)\otimes\chi(\lambda).$$

*Proof.* Note that, since[6]
$$\Lambda^n = (\delta_{i,j-n})_{i,j\in\mathbb{Z}}, \quad \varepsilon^n = ((i)_n\delta_{i,j+n})_{i,j\in\mathbb{Z}},$$
$$\Lambda^{*n} = (\delta_{i,j+n})_{i,j\in\mathbb{Z}}, \quad \varepsilon^{*n} = ((-i)_n\delta_{i,j-n})_{i,j\in\mathbb{Z}},$$

we have
$$\delta(\lambda,\Lambda) = \sum_{n\in\mathbb{Z}}\lambda^n\Lambda^{-n-1} = \sum_{n\in\mathbb{Z}}\lambda^n(\delta_{i,j+n+1})_{i,j\in\mathbb{Z}} = (\lambda^{i-j-1})_{i,j\in\mathbb{Z}},$$
$$\delta(\lambda,\Lambda^*) = \sum_{n\in\mathbb{Z}}\lambda^n\Lambda^{n+1} = \sum_{n\in\mathbb{Z}}\lambda^n(\delta_{i,j-n-1})_{i,j\in\mathbb{Z}} = (\lambda^{j-i-1})_{i,j\in\mathbb{Z}},$$

and
$$e^{(\mu-\lambda)\varepsilon} = \sum_{n=0}^{\infty}\frac{(\mu-\lambda)^n}{n!}((i)_n\delta_{i,j+n})_{i,j\in\mathbb{Z}} = \left(\binom{i}{i-j}(\mu-\lambda)^{i-j}\right)_{i,j\in\mathbb{Z}},$$
$$e^{(\mu-\lambda)\varepsilon^*} = \sum_{n=0}^{\infty}\frac{(\mu-\lambda)^n}{n!}((-i)_n\delta_{i,j-n})_{i,j\in\mathbb{Z}} = \left(\binom{-i}{-i+j}(\mu-\lambda)^{j-i}\right)_{i,j\in\mathbb{Z}}.$$

Hence
$$e^{(\mu-\lambda)\varepsilon}\delta(\lambda,\Lambda) = \left(\sum_{\substack{k\\i-k\geq 0}}\binom{i}{i-k}(\mu-\lambda)^{i-k}\lambda^{k-j-1}\right)_{i,j\in\mathbb{Z}}$$
$$= (\mu^i\lambda^{-j-1})_{i,j\in\mathbb{Z}},$$
$$e^{(\mu-\lambda)\varepsilon^*}\delta(\lambda,\Lambda^*) = \left(\sum_{\substack{k\\k-i\geq 0}}\binom{-i}{-i+k}(\mu-\lambda)^{k-i}\lambda^{-k+j-1}\right)_{i,j\in\mathbb{Z}}$$
$$= (\mu^{-i}\lambda^{j-1})_{i,j\in\mathbb{Z}}. \qquad\square$$

---

[5] Given two column vectors $a$ and $b$, the matrix $a\otimes b$ is defined componentwise as follows
$$(a\otimes b)_{ij} = a_i b_j.$$

[6] For any $k,n\in\mathbb{Z}$, $n\geq 0$, we use the standard notation $(k)_n := k(k-1)...(k-n+1)$ and $\binom{k}{n} = \frac{(k)_n}{n!}$.

## 2. The two-Toda lattice

Consider the splitting of the algebra $\mathcal{D}$ of pairs $(P_1, P_2)$ of infinite $(\mathbb{Z} \times \mathbb{Z})$ matrices such that $(P_1)_{ij} = 0$ for $j - i \gg 0$ and $(P_2)_{ij} = 0$ for $i - j \gg 0$, used in [6]; to wit:

$$\begin{aligned}
\mathcal{D} &= \mathcal{D}_+ + \mathcal{D}_-, \\
\mathcal{D}_+ &= \left\{ (P, P) \mid P_{ij} = 0 \text{ if } |i-j| \gg 0 \right\} = \left\{ (P_1, P_2) \in \mathcal{D} \mid P_1 = P_2 \right\}, \\
\mathcal{D}_- &= \left\{ (P_1, P_2) \mid (P_1)_{ij} = 0 \text{ if } j \geq i, \ (P_2)_{ij} = 0 \text{ if } i > j \right\},
\end{aligned}$$

with $(P_1, P_2) = (P_1, P_2)_+ + (P_1, P_2)_-$ given by

(2.1) $$\begin{aligned}
(P_1, P_2)_+ &= (P_{1u} + P_{2\ell}, P_{1u} + P_{2\ell}), \\
(P_1, P_2)_- &= (P_{1\ell} - P_{2\ell}, P_{2u} - P_{1u});
\end{aligned}$$

$P_u$ and $P_\ell$ denote the upper (including diagonal) and strictly lower triangular parts of the matrix $P$, respectively.

Throughout this paper, we will use the following operators $e^{\xi_i(z)}$ (a multiplication operator) and $e^{\eta_i(z)}$ (a shift), where

(2.2) $$\xi_1(z) = \sum_1^\infty t_k z^k \quad \text{and} \quad \xi_2(z) = \sum_1^\infty s_k z^{-k},$$

$$\eta_1(z) = \sum_1^\infty \frac{z^{-i}}{i} \frac{\partial}{\partial t_i} \quad \text{and} \quad \eta_2(z) = \sum_1^\infty \frac{z^i}{i} \frac{\partial}{\partial s_i},$$

so that

$$e^{a\eta_1 + b\eta_2} f(t, s) = f(t + a[z^{-1}], s + b[z])$$

with $[\alpha] = (\alpha, \alpha^2/2, \alpha^3/3, \ldots)$.

The two-dimensional Toda lattice equations

(2.3) $$\frac{\partial L}{\partial t_n} = \left[ \left( L_1^n, 0 \right)_+, L \right] \quad \text{and} \quad \frac{\partial L}{\partial s_n} = \left[ \left( 0, L_2^n \right)_+, L \right] \quad , \ n = 1, 2, \ldots$$

are deformations of a pair of infinite matrices

(2.4) $$L = (L_1, L_2) = \left( \sum_{-\infty < i \leq 1} a_i^{(1)} \Lambda^i, \sum_{-1 \leq i < \infty} a_i^{(2)} \Lambda^i \right) \in \mathcal{D},$$

with $\Lambda$ the shift operator of Section 1 and where $a_i^{(1)}$ and $a_i^{(2)}$ are diagonal matrices depending on $t = (t_1, t_2, \ldots)$ and $s = (s_1, s_2, \ldots)$, such that

$$a_1^{(1)} = I \quad \text{and} \quad \left( a_{-1}^{(2)} \right)_{nn} \neq 0 \quad \text{for all } n.$$

In analogy with Sato's theory, in [18] it is shown that a solution $L$ of (2.3) has the representation

$$L_1 = W_1 \Lambda W_1^{-1} = S_1 \Lambda S_1^{-1}, \quad L_2 = W_2 \Lambda^{-1} W_2^{-1} = S_2 \Lambda^{-1} S_2^{-1}$$

in terms of two pairs of wave operators

$$\begin{cases} S_1 = \sum_{i \leq 0} c_i(t,s)\Lambda^i, \quad S_2 = \sum_{i \geq 0} c'_i(t,s)\Lambda^i \\ c_i, c'_i : \text{diagonal matrices}, c_0 = I, (c'_0)_{ii} \neq 0, \text{ for all } i \end{cases}$$

and

(2.5) $$W_i = S_i(t,s)e^{\xi_i(\Lambda)}.$$

One also introduces pairs of wave and adjoint wave vectors $\Psi = (\Psi_1, \Psi_2)$, and $\Psi^* = (\Psi_1^*, \Psi_2^*)$:

(2.6) $$\begin{aligned} \Psi_i(t,s;z) &= W_i\chi(z) = e^{\xi_i(z)}S_i\chi(z), \\ \Psi_i^*(t,s;z) &= (W_i^\top)^{-1}\chi^*(z) = e^{-\xi_i(z)}(S_i^\top)^{-1}\chi^*(z), \end{aligned}$$

which evolve in $t$ and $s$ according to the following differential equations:[7]

(2.7) $$\begin{cases} \frac{\partial}{\partial t_n}\Psi = (L_1^n, 0)_+ \Psi = ((L_1^n)_u, (L_1^n)_u)\Psi \\ \frac{\partial}{\partial s_n}\Psi = (0, L_2^n)_+ \Psi = ((L_2^n)_\ell, (L_2^n)_\ell)\Psi \end{cases}$$

$$\begin{cases} \frac{\partial}{\partial t_n}\Psi^* = -((L_1^n, 0)_+)^\top \Psi^* \\ \frac{\partial}{\partial s_n}\Psi^* = -((0, L_2^n)_+)^\top \Psi^*. \end{cases}$$

Besides $L = (L_1, L_2)$, we define the operators $L^* = (L_1^*, L_2^*)$, $M = (M_1, M_2)$ and $M^* = (M_1^*, M_2^*)$ as follows

(2.8)
$$L := (W_1 \Lambda W_1^{-1}, W_2 \Lambda^* W_2^{-1}), \quad L^* := ((W_1^\top)^{-1}\Lambda^* W_1^\top, (W_2^\top)^{-1}\Lambda W_2^\top),$$
$$M := (W_1 \varepsilon W_1^{-1}, W_2 \varepsilon^* W_2^{-1}), \quad M^* := ((W_1^\top)^{-1}\varepsilon^* W_1^\top, (W_2^\top)^{-1}\varepsilon W_2^\top),$$

which satisfy, in view of (2.6) and (1.1):

$$L\Psi = (z, z^{-1})\Psi, \quad M\Psi = \left(\frac{\partial}{\partial z}, \frac{\partial}{\partial (z^{-1})}\right)\Psi, \quad [L, M] = (1, 1),$$

$$L^*\Psi^* = (z, z^{-1})\Psi^*, \quad M^*\Psi = \left(\frac{\partial}{\partial z}, \frac{\partial}{\partial (z^{-1})}\right)\Psi^*, \quad [L^*, M^*] = (1, 1).$$

---

[7]Here the action is viewed componentwise, e.g., $(A, B)\Psi = (A\Psi_1, B\Psi_2)$ or $(z, z^{-1})\Psi = (z\Psi_1, z^{-1}\Psi_2)$.

The operators $L, M, L^\top, M^\top$ and $W := (W_1, W_2)$ evolve according to

(2.9)
$$\begin{cases} \frac{\partial}{\partial t_n}\begin{pmatrix} L \\ M \end{pmatrix} = \left[(L_1^n, 0)_+, \begin{pmatrix} L \\ M \end{pmatrix}\right], \\ \frac{\partial}{\partial s_n}\begin{pmatrix} L \\ M \end{pmatrix} = \left[(0, L_2^n)_+, \begin{pmatrix} L \\ M \end{pmatrix}\right] \end{cases}$$

$$\frac{\partial W}{\partial t_n} = (L_1^n, 0)_+ W, \quad \text{and} \quad \frac{\partial W}{\partial s_n} = (0, L_2^n)_+ W.$$

Ueno and Takasaki [18] show that the two-Toda deformations of $\Psi$, and hence $L$, can ultimately all be expressed in terms of one sequence of $\tau$-functions

$$\tau(n,t,s) = \tau_n(t_1, t_2, \ldots; s_1, s_2, \ldots) = \det[(S_1^{-1} S_2(t,s))_{i,j}]_{-\infty \leq i,j \leq n-1}, \quad n \in \mathbb{Z}:$$

to wit:

(2.10)
$$\Psi_1(t,s;z) = \left(\frac{e^{-\eta_1} \tau_n(t,s)}{\tau_n(t,s)} e^{\sum_1^\infty t_i z^i} z^n\right)_{n \in \mathbb{Z}}$$

$$\Psi_2(t,s;z) = \left(\frac{e^{-\eta_2} \tau_{n+1}(t,s)}{\tau_n(t,s)} e^{\sum_1^\infty s_i z^{-i}} z^n\right)_{n \in \mathbb{Z}}$$

$$\Psi_1^*(t,s;z) = \left(\frac{e^{\eta_1} \tau_{n+1}(t,s)}{\tau_{n+1}(t,s)} e^{-\sum_1^\infty t_i z^i} z^{-n}\right)_{n \in \mathbb{Z}}$$

$$\Psi_2^*(t,s;z) = \left(\frac{e^{\eta_2} \tau_n(t,s)}{\tau_{n+1}(t,s)} e^{-\sum_1^\infty s_i z^{-i}} z^{-n}\right)_{n \in \mathbb{Z}}.$$

Finally the pair of matrices $W = (W_1, W_2)$ satisfies the bilinear relation (in the $\pm$ splitting of (2.1))

$$(W(t,s) W(t', s')^{-1})_- = 0$$

or equivalently,

(2.11) $$W_1(t,s) W_1(t', s')^{-1} = W_2(t,s) W_2(t', s')^{-1},$$

from which one proves Proposition 2.1; for details see [6]. Equation (2.13) below is established in Adler-van Moerbeke [4].

PROPOSITION 2.1 (bi-infinite and semi-infinite). *The wave and adjoint wave functions satisfy, for all $m, n \in \mathbb{Z}$ (bi-infinite) and $m, n \geq 0$ (semi-infinite) and $t, s, t', s' \in \mathbb{C}^\infty$:*

(2.12)
$$\oint_{z=\infty} \Psi_{1n}(t,s;z) \Psi_{1m}^*(t',s';z) \frac{dz}{2\pi i z} = \oint_{z=0} \Psi_{2n}(t,s;z) \Psi_{2m}^*(t',s';z') \frac{dz}{2\pi i z}.$$

*Moreover $L_1$ has the following representation in terms of $\tau$-functions*

$$(2.13) \qquad L_1^k = \sum_{\ell=0}^{\infty} \operatorname{diag} \left( \frac{p_\ell(\tilde{\partial}/\partial t)\tau_{n+k-\ell+1} \circ \tau_n}{\tau_{n+k-\ell+1}\tau_n} \right)_{n\in\mathbb{Z}} \Lambda^{k-\ell}.$$

In the semi-infinite case, we reinterpret $\Lambda^{-1}$ as $\Lambda^\top$, where $\Lambda$ is the semi-infinite shift operator. Then one shows (2.11) and (2.12) are also valid. Also the semi-infinite case is obtained from the infinite case by setting $\tau_{-i} = 0$ for $i \geq 1$ and $\tau_0 = 1$. Then the semi-infinite wave vectors

$$\left(\Psi_{1n}(t,s;z)e^{-\sum t_i z^i}\right)_{n\geq 0} \quad \text{and} \quad \left(\Psi_{2n}^*(t,s;z)e^{\sum s_i z^{-i}}\right)_{n\geq 0}$$

are vectors of polynomials of degree $n = 0, 1, 2, \ldots$ in $z$ and $z^{-1}$ respectively, as follows from (2.12); see [2]. In the semi-infinite case, we must define $L_1^n$ and $L_2^{\top n}$ for integers $n < 0$; namely for $n \in \mathbb{Z}$:

$$(2.14) \qquad \begin{aligned} L_1^{(n)}(e^{-\sum t_i z^i}\Psi_1(z)) &= \pi_+(z^n e^{-\sum t_i z^i}\Psi_1(z)), \\ L_2^{\top(n)}(e^{\sum s_i z^i}\Psi_2^*(z^{-1})) &= \pi_+(z^n e^{\sum s_i z^i}\Psi_2^*(z^{-1})), \end{aligned}$$

where $\pi_+$ refers to the projection $\pi_+(\sum_{i\in\mathbb{Z}} a_i z^i) = \sum_{i\geq 0} a_i z^i$. Observe that $L_1^{(n)} = L_1^n$, $L_2^{\top(n)} = L_2^{\top n}$ for $n \in \mathbb{Z}$, $n \geq 0$; indeed, multiplying the vector $\Psi_1(z)e^{-\Sigma t_i z^i}$ of polynomials with $z^n$, $n \geq 0$ maintains, the polynomial character, and thus for $n \geq 0$,

$$L_1^{(n)}(\Psi_1(z)) = \pi_+(z^n \Psi_1(z)) = z^n \Psi_1(z) = L_1^n \Psi(z).$$

## 3. Bilinear Fay identities and a new identity for two-Toda $\tau$-functions

Two-Toda $\tau$-functions $\tau(t,s)$ satisfy the KP-hierarchy in $t$ and $s$ separately, of which the first equation reads:

$$\left(\frac{\partial}{\partial t_1}\right)^4 \log \tau + 6\left(\left(\frac{\partial}{\partial t_1}\right)^2 \log \tau\right)^2 + 3\left(\frac{\partial}{\partial t_2}\right)^2 \log \tau - 4\frac{\partial^2}{\partial t_1 \partial t_3}\log \tau = 0.$$

But they also satisfy the following identity:

THEOREM 3.1. *Two-Toda $\tau$-functions satisfy:*[8]

$$(3.1) \qquad \left\{\frac{\partial^2 \log \tau_n}{\partial t_1 \partial s_2}, \frac{\partial^2 \log \tau_n}{\partial t_1 \partial s_1}\right\}_{t_1} + \left\{\frac{\partial^2 \log \tau_n}{\partial s_1 \partial t_2}, \frac{\partial^2 \log \tau_n}{\partial t_1 \partial s_1}\right\}_{s_1} = 0.$$

---

[8]in terms of the Wronskian $\{f,g\}_t = \frac{\partial f}{\partial t}g - f\frac{\partial g}{\partial t}$.

The proof of this theorem hinges on the bilinear identity, due to Ueno-Takasaki [18] and a number of lemmas:

PROPOSITION 3.2. *Two-Toda $\tau$-functions satisfy the following bilinear identities*:

$$
\oint_{z=\infty} \tau_n(t-[z^{-1}],s)\tau_{m+1}(t'+[z^{-1}],s')e^{\sum_1^\infty (t_i-t'_i)z^i} z^{n-m-1} dz
$$
$$
= \oint_{z=0} \tau_{n+1}(t,s-[z])\tau_m(t',s'+[z])e^{\sum_1^\infty (s_i-s'_i)z^{-i}} z^{n-m-1} dz, \quad (3.2)
$$

or, expressed in terms of the Hirota symbol,[9]

$$
\sum_{j=0}^\infty p_{m-n+j}(-2a)p_j(\tilde{\partial}_t)e^{\sum_1^\infty (a_k \frac{\partial}{\partial t_k}+b_k \frac{\partial}{\partial s_k})}\tau_{m+1}\circ \tau_n \quad (3.3)
$$
$$
= \sum_{j=0}^\infty p_{-m+n+j}(-2b)p_j(\tilde{\partial}_s)e^{\sum_1^\infty (a_k \frac{\partial}{\partial t_k}+b_k \frac{\partial}{\partial s_k})}\tau_m\circ \tau_{n+1},
$$

*both, for the bi-infinite ($n,m \in \mathbb{Z}$) and the semi-infinite case ($n,m \in \mathbb{Z}$, $n,m \geq 0$).*

*Proof.* (3.2) follows at once from Proposition 2.1 and the $\tau$-function representations (2.10), whereas (3.3) follows from the shifts $t \mapsto t - a$, $t' \mapsto t' + a$, $s \mapsto s - b$, $s' \mapsto s' + b$, combined with the definition of the Hirota symbol. □

PROPOSITION 3.3.

$$
\left(L_1^k\right)_{n,n+1} = \frac{p_{k-1}(\tilde{\partial}_t)\tau_{n+2}\circ \tau_n}{\tau_{n+2}\tau_n} = \frac{\frac{\partial^2 \log \tau_{n+1}}{\partial s_1 \partial t_k}}{\frac{\partial^2 \log \tau_{n+1}}{\partial s_1 \partial t_1}} \quad (3.4)
$$

$$
\left(hL_2^{\top k}h^{-1}\right)_{n,n+1} = \frac{p_{k-1}(-\tilde{\partial}_s)\tau_{n+2}\circ \tau_n}{\tau_{n+2}\tau_n} = \frac{\frac{\partial^2 \log \tau_{n+1}}{\partial t_1 \partial s_k}}{\frac{\partial^2 \log \tau_{n+1}}{\partial s_1 \partial t_1}}. \quad (3.5)
$$

*Proof.* Set $m = n+1$, all $b_k$ and $a_k = 0$, except for one $a_{j+1}$, in the Hirota bilinear relation (3.3). The first nonzero term in the sum on the left-hand side of that relation, which is also the only one containing $a_{j+1}$ linearly, reads
$$
p_{j+1}(-2a)p_j(\tilde{\partial}_t)e^{a_{j+1}\frac{\partial}{\partial t_{j+1}}}\tau_{n+2}\circ \tau_n + ... = -2a_{j+1}p_j(\tilde{\partial}_t)\tau_{n+2}\circ \tau_n + O(a_{j+1}^2), \quad (3.6)
$$

whereas the right-hand side equals

$$
p_0(0)p_1(\tilde{\partial}_s)e^{a_{j+1}\frac{\partial}{\partial t_{j+1}}}\tau_{n+1}\circ \tau_{n+1} = \frac{\partial}{\partial s_1}(1 + a_{j+1}\frac{\partial}{\partial t_{j+1}} + ...)\tau_{n+1}\circ \tau_{n+1}. \quad (3.7)
$$

---

[9]For the customary Hirota symbol $p(\partial_t)f \circ g := p(\frac{\partial}{\partial y})f(t+y)g(t-y)\big|_{y=0}$.

Comparing the coefficients of $a_{j+1}$ in (3.4) and (3.5) yields

$$-2\, p_j(\tilde{\partial}_t)\tau_{n+2}\circ\tau_n = \frac{\partial^2}{\partial s_1 \partial t_{j+1}}\tau_{n+1}\circ\tau_{n+1};$$

in particular, we have

(3.8) $$\frac{p_{k-1}(\tilde{\partial}_t)\tau_{n+2}\circ\tau_n}{\tau_{n+1}^2} = -\frac{\partial^2}{\partial s_1 \partial t_k}\log\tau_{n+1},$$

and so, for $k=1$,

(3.9) $$\frac{\tau_n\tau_{n+2}}{\tau_{n+1}^2} = -\frac{\partial^2}{\partial s_1 \partial t_1}\log\tau_{n+1}.$$

Dividing (3.8) and (3.9) leads to the second equality in (3.4). But, according to (2.13), the $(n,n+1)$-entry of $L_1^k$ is given by (3.4). The similar result for $L_2^k$ is given by the involution

$$t \longleftrightarrow -s \text{ and } L_1 \longleftrightarrow hL_2^\top h^{-1}. \qquad \square$$

*Proof of Theorem 3.1.* Set $k=2$ in the identities of Proposition 3.3; then subtracting $\frac{\partial}{\partial t_1}$ of identity (3.5) from $\frac{\partial}{\partial s_1}$ of (3.4) leads to Theorem 3.1. $\square$

The $(n,n+1)$-entries of $L_1^2$ and $hL_2^{\top 2}h^{-1}$ have the following equivalent expressions, which will be useful in the theory of Toeplitz matrices, as applied to the distribution of the length of the longest increasing sequences of a random permutation. The second identity, appearing in $L_1^2$ below is an expression purely in terms of one component $\tau_{n+1}$, at the expense of introducing a $\partial/\partial t_2$-derivative; the third identity involves $\partial/\partial t_1$ and $\partial/\partial s_1$ only, but at the expense of involving nearest neighbors $\tau_n$ and $\tau_{n+1}$.

LEMMA 3.4. *Two-Toda $\tau$-functions satisfy*:

(3.10) $$\left(L_1^2\right)_{n,n+1} = \frac{\partial}{\partial t_1}\log\frac{\tau_{n+2}}{\tau_n}$$

$$= \frac{\frac{\partial^2}{\partial s_1 \partial t_2}\log\tau_{n+1}}{\frac{\partial^2}{\partial s_1 \partial t_1}\log\tau_{n+1}}$$

$$= \frac{\partial}{\partial t_1}\log\left(\left(\frac{\tau_{n+1}}{\tau_n}\right)^2 \frac{\partial^2}{\partial s_1 \partial t_1}\log\tau_{n+1}\right)$$

$$\left(hL_2^{\top 2}h^{-1}\right)_{n,n+1} = -\frac{\partial}{\partial s_1}\log\frac{\tau_{n+2}}{\tau_n}$$

$$= \frac{\frac{\partial^2}{\partial t_1 \partial s_2}\log\tau_{n+1}}{\frac{\partial^2}{\partial t_1 \partial s_1}\log\tau_{n+1}}$$

$$= -\frac{\partial}{\partial s_1}\log\left(\left(\frac{\tau_{n+1}}{\tau_n}\right)^2 \frac{\partial^2}{\partial s_1 \partial t_1}\log\tau_{n+1}\right).$$

*Proof.* From Proposition 3.3 ($k=2$), it follows that

$$\frac{\partial^2 \log \tau_{n+1}}{\partial s_1 \partial t_2} = \frac{\partial^2 \log \tau_{n+1}}{\partial s_1 \partial t_1} \frac{\frac{\partial}{\partial t_1}(\tau_{n+2} \circ \tau_n)}{\tau_{n+2}\tau_n}$$

$$= \frac{\partial^2 \log \tau_{n+1}}{\partial s_1 \partial t_1} \frac{\partial}{\partial t_1} \log \frac{\tau_{n+2}}{\tau_n}$$

$$= \frac{\partial^2 \log \tau_{n+1}}{\partial s_1 \partial t_1} \left( \frac{\partial}{\partial t_1} \log \frac{\tau_{n+2}}{\tau_{n+1}} + \frac{\partial}{\partial t_1} \log \frac{\tau_{n+1}}{\tau_n} \right)$$

$$= \frac{\partial^2 \log \tau_{n+1}}{\partial s_1 \partial t_1} \left( \frac{\partial}{\partial t_1} \log \left( -\frac{\tau_{n+1}}{\tau_n} \frac{\partial^2}{\partial s_1 \partial t_1} \log \tau_{n+1} \right) \right.$$

$$\left. + \frac{\partial}{\partial t_1} \log \frac{\tau_{n+1}}{\tau_n} \right), \text{ using (3.9)}$$

$$= \frac{\partial^2 \log \tau_{n+1}}{\partial s_1 \partial t_1} \left( 2\frac{\partial}{\partial t_1} \log \frac{\tau_{n+1}}{\tau_n} + \frac{\partial}{\partial t_1} \log \left( -\frac{\partial^2}{\partial s_1 \partial t_1} \log \tau_{n+1} \right) \right)$$

$$= 2\frac{\partial}{\partial t_1} \log \frac{\tau_{n+1}}{\tau_n} \frac{\partial^2}{\partial s_1 \partial t_1} \log \tau_{n+1} + \frac{\partial}{\partial t_1} \left( \frac{\partial^2}{\partial s_1 \partial t_1} \log \tau_{n+1} \right).$$

The second to the last equation establishes the first equation (3.10). The second equation (3.10) is simply the dual of the first one by $t_i \leftrightarrow -s_i$. □

The remaining statements in this section hold for both the bi-infinite case and the semi-infinite case; we thank T. Shiota for showing us how shifting the arguments in various directions, and repeatedly, leads to many different identities.

PROPOSITION 3.5 (the Fay identity). *If*

$$(3.11) \quad \alpha - \alpha' = \sum_1^p [z_k^{-1}] - \sum_1^q [y_k^{-1}] \text{ and } \beta - \beta' = \sum_1^{p'} [v_k] - \sum_1^{q'} [u_k],$$

*with $p, q, p', q' \geq 0$, we have*

$$(3.12) \quad \sum_{\ell=1}^p \tau_n(\alpha - [z_\ell^{-1}], \beta)\tau_{m+1}(\alpha' + [z_\ell^{-1}], \beta') z_\ell^{r+1} \frac{\prod_{k=1}^q (z_\ell^{-1} - y_k^{-1})}{\prod_{\substack{k=1 \\ k \neq \ell}}^p (z_\ell^{-1} - z_k^{-1})}$$

$$+ \frac{1}{r!} \frac{\partial^r}{\partial x^r} \left( \tau_n(\alpha - [x], \beta)\tau_{m+1}(\alpha' + [x], \beta') \frac{\prod_1^q (x - y_k^{-1})}{\prod_1^p (x - z_k^{-1})} \right) \bigg|_{x=0}$$

$$= \sum_{\ell=1}^{p'} \tau_{n+1}(\alpha, \beta - [v_\ell])\tau_m(\alpha', \beta' + [v_\ell])v_\ell^{-r'-1}\frac{\prod_{k=1}^{q'}(v_\ell - u_k)}{\prod_{\substack{k=1\\k\neq\ell}}^{p'}(v_\ell - v_k)}$$

$$+ \frac{1}{r'!}\frac{\partial^{r'}}{\partial x^{r'}}\left(\tau_{n+1}(\alpha, \beta - [x])\tau_m(\alpha', \beta' + [x])\frac{\prod_1^{q'}(x - u_k)}{\prod_1^{p'}(x - v_k)}\right)\bigg|_{x=0},$$

where $r := n - m + q - p$ and $r' := -n + m + q' - p'$, with the understanding that $(\partial/\partial x)^r = 0$ for $r < 0$.

*Proof.* The relations (3.11) imply

$$(3.13) \qquad e^{\Sigma_1^\infty (\alpha_i - \alpha'_i)z^i} = \frac{\prod_1^q\left(1 - \frac{z}{y_k}\right)}{\prod_1^p\left(1 - \frac{z}{z_k}\right)} \text{ and } e^{\Sigma_1^\infty (\beta_i - \beta'_i)z^{-i}} = \frac{\prod_1^{q'}\left(1 - \frac{u_k}{z}\right)}{\prod_1^{p'}\left(1 - \frac{v_k}{z}\right)}.$$

If $f$ denotes a holomorphic function in a large enough disc around $z = \infty$, as in

$$f(z) := \tau_n(\alpha - [z^{-1}], \beta)\tau_{m+1}(\alpha' + [z^{-1}], \beta'),$$

then we have, using (3.13),

$$\frac{1}{2\pi i}\oint_{z=\infty} f(z)e^{\Sigma_1^\infty(\alpha_i - \alpha'_i)z^i}z^{n-m-1}dz$$

$$= \frac{1}{2\pi i}\oint_{z=\infty} f(z)\frac{\prod_1^q\left(1 - \frac{z}{y_k}\right)}{\prod_1^p\left(1 - \frac{z}{z_k}\right)}z^{n-m-1}dz$$

$$\overset{*}{=} \frac{1}{2\pi i}\oint_{x=0} f(x^{-1})\frac{\prod_1^q(x - y_k^{-1})}{\prod_1^p(x - z_k^{-1})}x^{-n+m-q+p-1}dx, \text{ upon setting } x = \tfrac{1}{z}$$

$$= \sum_{\ell=1}^p f(z_\ell)\frac{\prod_{k=1}^q(z_\ell^{-1} - y_k^{-1})}{\prod_{\substack{k=1\\k\neq\ell}}^p(z_\ell^{-1} - z_k^{-1})}z_\ell^{n-m+q-p+1}$$

$$+ \frac{1}{r!}\left(\frac{d}{dx}\right)^r f(x^{-1})\frac{\prod_1^q(x - y_k^{-1})}{\prod_1^p(x - z_k^{-1})}\bigg|_{x=0},$$

which uses the fact that the integrand has poles at $x = z_k^{-1}$ $(1 \leq k \leq p)$ and at $x = 0$, if $r = n - m + q - p \geq 0$. In $\overset{*}{=}$ the sign change in $dx$ due to $x = 1/z$ and the change of orientation of the contour integration cancel each other out.

If $f'$ denotes a holomorphic function in a large enough disc around $z = 0$, as in

$$f'(z) := \tau_{n+1}(\alpha, \beta - [z])\tau_m(\alpha', \beta' + [z]),$$

then we have, again using (3.13),

$$\frac{1}{2\pi i} \oint_{z=0} f'(z) e^{\Sigma(\beta_i - \beta'_i)z^{-i}} z^{n-m-1} dz$$

$$= \frac{1}{2\pi i} \oint_{z=0} f'(z) \frac{\prod_1^{q'}(1 - \frac{u_k}{z})}{\prod_1^{p'}(1 - \frac{v_k}{z})} z^{n-m-1} dz$$

$$= \frac{1}{2\pi i} \oint_{z=0} f'(z) \frac{\prod_1^{q'}(z - u_k)}{\prod_1^{p'}(z - v_k)} z^{n-m+p'-q'-1} dz$$

$$= \sum_{\ell=1}^{p'} f'(v_\ell) \frac{\prod_{k=1}^{q'}(v_\ell - u_k)}{\prod_{\substack{k=1\\k\neq\ell}}^{p'}(v_\ell - v_k)} v_\ell^{n-m+p'-q'-1}$$

$$+ \frac{1}{r'!} \left(\frac{d}{dx}\right)^{r'} f'(x) \frac{\prod_1^{q'}(x - u_k)}{\prod_1^{p'}(x - v_k)} \bigg|_{z=0},$$

which used the fact that the integrand has poles at $z = v_k$ ($1 \leq k \leq p'$) and at $z = 0$, if $r' = -n + m - p' + q' \geq 0$. $\square$

COROLLARY 3.6.

$$\tau_n(t - [z^{-1}], s + [v] - [u])\tau_n(t, s) - \tau_n(t, s + [v] - [u])\tau_n(t - [z^{-1}], s)$$
$$= \frac{v - u}{z} \tau_{n+1}(t, s - [u])\tau_{n-1}(t - [z^{-1}], s + [v]).$$

*Proof.* Setting $m = n - 1$, $\alpha = t$, $\beta = s + [v] - [u]$, $\alpha' = t - [z^{-1}]$, $\beta' = s$, we have $\alpha - \alpha' = [z^{-1}]$ and $\beta - \beta' = [v] - [u]$, and thus $p = p' = q' = 1$, $q = 0$, with $0 = -p' + q' < n - m = p - q = 1$; that is, $r = 0$, $r' = -1$. Then Proposition 3.5 leads to the proof of Corollary 3.6. $\square$

COROLLARY 3.7.

$$\tau_n(t, s + [v_1])\tau_{n+1}(t + [z_1^{-1}] - [z_2^{-1}], s - [v_2]) \frac{z_1^{-1}}{z_1^{-1} - z_2^{-1}}$$
$$+ \tau_n(t + [z_1^{-1}] - [z_2^{-1}], s + [v_1])\tau_{n+1}(t, s - [v_2]) \frac{z_2^{-1}}{z_2^{-1} - z_1^{-1}}$$
$$= \tau_{n+1}(t + [z_1^{-1}], s)\tau_n(t - [z_2^{-1}], s + [v_1] - [v_2]) \frac{v_1}{v_1 - v_2}$$
$$+ \tau_{n+1}(t + [z_1^{-1}], s + [v_1] - [v_2])\tau_n(t - [z_2^{-1}], s) \frac{v_2}{v_2 - v_1}.$$

*Proof.* Setting $m = n$, $\alpha = t + [z_1^{-1}]$, $\alpha' = t - [z_2^{-1}]$, $\beta = s + [v_1]$, $\beta' = s - [v_2]$, we have $\alpha - \alpha' = [z_1^{-1}] + [z_2^{-1}]$ and $\beta - \beta' = [v_1] + [v_2]$, and thus $p = 2$, $q = 0$, $p' = 2$, $q' = 0$, with $-2 = -p' + q' < 0 = n - m < p - q = 2$, and so, $r = r' = -2$. Similarly Proposition 3.5 ends the proof of Corollary 3.7. $\square$

## 4. Higher Fay identities for the two-Toda lattice

LEMMA 4.1.

$$\sum_{\ell=1}^{k+1} \tau_{N-1}(t - [z_\ell^{-1}], s + [y_m^{-1}])\tau_{N-k}$$

$$\cdot \left(t - \sum_{\substack{j=1 \\ j \neq \ell}}^{k+1}[z_j^{-1}], s + \sum_{\substack{i=1 \\ i \neq m}}^{k+1}[y_i^{-1}]\right) \frac{1}{\prod_{\substack{i=1 \\ i \neq \ell}}^{k+1}(z_\ell^{-1} - z_i^{-1})}$$

$$= \tau_N(t,s)\tau_{N-k-1}\left(t - \sum_1^{k+1}[z_i^{-1}], s + \sum_1^{k+1}[y_i^{-1}]\right) \prod_{\substack{i=1 \\ i \neq m}}^{k}(y_m^{-1} - y_i^{-1}).$$

*Proof.* In Proposition 3.5, we set

$$\alpha = t, \qquad \alpha' = t - \sum_{j=1}^{k+1}[z_j^{-1}],$$

$$\beta = s + [y_m^{-1}], \qquad \beta' = s + \sum_{\substack{i=1 \\ i \neq m}}^{k+1}[y_i^{-1}].$$

Now obviously,

$$\alpha - \alpha' = \sum_{j=1}^{k+1}[z_j^{-1}] \text{ and } \beta - \beta' = [y_m^{-1}] - \sum_{\substack{i=1 \\ i \neq m}}^{k+1}[y_i^{-1}],$$

$$p = k+1, q = 0, p' = 1, q' = k, \quad n = N-1, \quad m = N-k-1, \quad r = r' = -1.$$

With these data, we have

$$\sum_{\ell=1}^{k+1} \tau_{N-1}(\alpha - [z_\ell^{-1}], \beta)\tau_{N-k}(\alpha' + [z_\ell^{-1}], \beta')\frac{1}{\prod_{\substack{i=1 \\ i \neq \ell}}^{k+1}(z_\ell^{-1} - z_i^{-1})}$$

$$= \tau_N(\alpha, \beta - [y_m^{-1}])\tau_{N-k-1}(\alpha', \beta' + [y_m^{-1}]) \prod_{\substack{i=1 \\ i \neq m}}^{k+1}(y_m^{-1} - y_i^{-1}),$$

establishing Lemma 4.1. □

THEOREM 4.2. *The two-Toda tau-functions $\tau(t,s)$ satisfy the following two higher Fay identities*:

$$(4.1) \quad \det\left(\frac{\tau_N(t-[z_i^{-1}]+[y_j^{-1}],s)}{\tau_N(t,s)}\frac{1}{y_j-z_i}\right)_{1\le i,j\le k}$$

$$= (-1)^{\frac{k(k-1)}{2}}\frac{\Delta(y)\Delta(z)}{\prod_{k,\ell}(y_k-z_\ell)}\frac{\tau_N\left(t+\sum_1^k[y_i^{-1}]-\sum_1^k[z_j^{-1}],s\right)}{\tau_N(t,s)};$$

$$\det\left(\frac{\tau_{N-1}(t-[z_i^{-1}],s+[y_j^{-1}])}{\tau_N(t,s)}\right)_{1\le i,j\le k}$$

$$= \Delta(y^{-1})\Delta(z^{-1})\frac{\tau_{N-k}\left(t-\sum_1^k[z_i^{-1}],s+\sum_1^k[y_i^{-1}]\right)}{\tau_N(t,s)}.$$

*Proof.* The inductive method for proving the first identity is due to [6]. As to the second relation, we also proceed by induction on the index $k$. Since the identity is obviously true for $k=1$, we assume it to be valid for $k\ge 1$ and we prove its validity for $k+1$. Indeed, by expanding the determinant according to the first column, we find

$$\det\left(\frac{\tau_{N-1}(t-[z_i^{-1}],s+[y_j^{-1}])}{\tau_N(t,s)}\right)_{1\le i,j\le k+1}$$

$$= \sum_{\ell=1}^{k+1}(-1)^{\ell-1}\frac{\tau_{N-1}(t-[z_\ell^{-1}],s+[y_1^{-1}])}{\tau_N(t,s)}\det\left(\frac{\tau_{N-1}(t-[z_i^{-1}],s+[y_j^{-1}])}{\tau_N(t,s)}\right)_{\substack{1\le i,j\le k+1\\i\ne\ell\\j\ne 1}}$$

$$= \sum_{\ell=1}^{k+1}(-1)^{\ell-1}\frac{\tau_{N-1}(t-[z_\ell^{-1}],s+[y_1^{-1}])}{\tau_N(t,s)}\prod_{\substack{1\le i<j\le k+1\\i\ne 1}}(y_i^{-1}-y_j^{-1})\prod_{\substack{1\le i<j\le k+1\\i,j\ne\ell}}(z_i^{-1}-z_j^{-1})$$

$$\cdot\frac{\tau_{N-k}\left(t-\sum_{i\ne\ell}[z_i^{-1}],s+\sum_{j\ne 1}[y_j^{-1}]\right)}{\tau_N(t,s)},\text{ by induction}$$

$$= \frac{\Delta(y^{-1})\Delta(z^{-1})}{\tau_N^2(t,s)}$$

$$\cdot\sum_{\ell=1}^{k+1}(-1)^{\ell-1}\frac{\tau_{N-1}(t-[z_\ell^{-1}],s+[y_1^{-1}])\tau_{N-k}\left(t-\sum_{\substack{i=1\\i\ne\ell}}^{k+1}[z_i^{-1}],s+\sum_{j=2}^{k+1}[y_j^{-1}]\right)}{\prod_{j=2}^{k+1}(y_1^{-1}-y_j^{-1})\prod_{\substack{j=1\\j\ne\ell}}^{k+1}(z_\ell^{-1}-z_j^{-1})(-1)^{\ell-1}}$$

$$= \Delta(y^{-1})\Delta(z^{-1})\frac{\tau_{N-k-1}\left(t-\sum_1^{k+1}[z_i^{-1}],s+\sum_1^{k+1}[y_i^{-1}]\right)}{\tau_N(t,s)},$$

by Lemma 4.1, ending the proof of Theorem 4.2. □

Such Fay identities were also obtained in the context of the multicomponent KP hierarchy by J. van de Leur [19].

## 5. Eigenfunction expansions and Vertex operators

In terms of the vertex operator,

$$X(t,\lambda) := e^{\sum_1^\infty t_i \lambda^i} e^{-\sum_1^\infty \lambda^{-i} \frac{1}{i} \frac{\partial}{\partial t_i}},$$

acting on functions $f(t_1, t_2, \ldots)$ of $t \in \mathbb{C}^\infty$ and using the diagonal matrix $\chi(\lambda)$, define the following four operators acting on column vectors $g = (g_n(t_1, t_2, \ldots))_{n \in \mathbb{Z}}$:

(5.1) $\quad \mathbb{X}_1(t,\mu) := X(t,\mu)\chi(\mu), \qquad \mathbb{X}_1^*(t,\lambda) := -\chi^*(\lambda) X(-t,\lambda),$

$\quad \mathbb{X}_2(s,\mu) := -X(s,\mu)\chi^*(\mu)\Lambda, \quad \mathbb{X}_2^*(s,\lambda) := \Lambda^{-1}\chi(\lambda) X(-s,\lambda),$

and the compositions
(5.2)
$$\begin{pmatrix} \mathbb{X}_{11}(\mu,\lambda) & \mathbb{X}_{21}(\mu,\lambda) \\ \mathbb{X}_{12}(\mu,\lambda) & \mathbb{X}_{22}(\mu,\lambda) \end{pmatrix} := (\mathbb{X}_1^*(t,\lambda) \ \mathbb{X}_2^*(s,\lambda)) \otimes (\mathbb{X}_1(t,\mu) \ \mathbb{X}_2(s,\mu))$$

$$= \begin{pmatrix} \mathbb{X}_1^*(t,\lambda)\mathbb{X}_1(t,\mu) & \mathbb{X}_1^*(t,\lambda)\mathbb{X}_2(s,\mu) \\ \mathbb{X}_2^*(s,\lambda)\mathbb{X}_1(t,\mu) & \mathbb{X}_2^*(s,\lambda)\mathbb{X}_2(s,\mu) \end{pmatrix}.$$

The main theorem of this section is the following[10]

THEOREM 5.1. *The following holds*: (i) *in the bi-infinite case*,

(5.3) $\begin{pmatrix} (\sum_{j<n} \Psi_{1j}^*(\lambda)\Psi_{1j}(\mu))_{n\in\mathbb{Z}} & (\sum_{j\geq n} \Psi_{1j}^*(\lambda)\Psi_{2j}(\mu^{-1}))_{n\in\mathbb{Z}} \\ (\sum_{j<n} \Psi_{2j}^*(\lambda^{-1})\Psi_{1j}(\mu))_{n\in\mathbb{Z}} & (\sum_{j\geq n} \Psi_{2j}^*(\lambda^{-1})\Psi_{2j}(\mu^{-1}))_{n\in\mathbb{Z}} \end{pmatrix}$

$= \dfrac{1}{\tau} \begin{pmatrix} \mathbb{X}_{11}(\mu,\lambda) & \mathbb{X}_{21}(\mu,\lambda) \\ \mathbb{X}_{12}(\mu,\lambda) & \mathbb{X}_{22}(\mu,\lambda) \end{pmatrix} \tau,$

*and* (ii) *in the semi-infinite case*,

(5.4) $\begin{pmatrix} (\sum_{0\leq j<n} \Psi_{1j}^*(\lambda)\Psi_{1j}(\mu))_{n>0} & (\sum_{j\geq n} \Psi_{1j}^*(\lambda)\Psi_{2j}(\mu^{-1}))_{n>0} \\ (\sum_{0\leq j<n} \Psi_{2j}^*(\lambda^{-1})\Psi_{1j}(\mu))_{n>0} & (\sum_{j\geq n} \Psi_{2j}^*(\lambda^{-1})\Psi_{2j}(\mu^{-1}))_{n>0} \end{pmatrix}$

$= \dfrac{1}{\tau}\left(\begin{pmatrix} \mathbb{X}_{11} & \mathbb{X}_{21} \\ \mathbb{X}_{12} & \mathbb{X}_{22} \end{pmatrix} + \begin{pmatrix} (1-\frac{\mu}{\lambda})^{-1} e^{\sum_1^\infty t_i(\mu^i - \lambda^i)} & 0 \\ 0 & 0 \end{pmatrix}\right)\tau.$

---

[10] For column vectors $v_1$, $v_2$, $w_1$ and $w_2$, we define

$$(v_1 \quad v_2) \otimes (w_1 \quad w_2) = \begin{pmatrix} v_1 \otimes w_1 & v_1 \otimes w_2 \\ v_2 \otimes w_1 & w_2 \otimes w_2 \end{pmatrix}.$$

COROLLARY 5.2 (eigenfunction expansion). *In the semi-infinite case, the functions below admit the following eigenfunction expansions*:

$$(5.5) \quad (1-\frac{\mu}{\lambda})^{-1} e^{\sum_1^\infty t_n(\mu^n - \lambda^n)} = \sum_{j=0}^\infty \Psi_{1j}^*(\lambda)\Psi_{1j}(\mu) \qquad |\mu| < |\lambda|$$

$$(1-\frac{\lambda}{\mu})^{-1} e^{\sum_1^\infty s_n(\mu^n - \lambda^n)} = \sum_{j=0}^\infty \Psi_{2j}^*(\lambda^{-1})\Psi_{2j}(\mu^{-1})$$

$$e^{\sum_1^\infty (s_n \mu^n - t_n \lambda^n)} \tau_1(t+[\lambda^{-1}], s-[\mu^{-1}]) = \sum_{j=0}^\infty \Psi_{1j}^*(\lambda)\Psi_{2j}(\mu^{-1}).$$

The proof of Theorem 5.1 and Corollary 5.2 relies on Lemmas 5.3 and 5.4.

*Remark* 1. With the identities

$$(5.6) \quad e^{-\sum_1^\infty a^i/i} = 1-a \quad \text{and} \quad \sum_0^\infty a^i = (1-a)^{-1},$$

the composition of $X(t,\mu)$ and $X(-t,\lambda)$ relates to the customary vertex operator $X(t,\lambda,\mu)$, as follows:

$$(5.7) \quad X(-t,\lambda)X(t,\mu)f = X(-t,\lambda)\left(e^{\sum_1^\infty t_i \mu^i} f(t-[\mu^{-1}])\right)$$

$$= e^{-\sum_1^\infty t_i \lambda^i} e^{\sum_1^\infty (t_i+\frac{\lambda^{-i}}{i})\mu^i} f(t+[\lambda^{-1}]-[\mu^{-1}])$$

$$= \frac{\lambda}{\lambda-\mu} e^{\Sigma t_i(\mu^i-\lambda^i)} f(t+[\lambda^{-1}]-[\mu^{-1}])$$

$$=: \frac{\lambda}{\lambda-\mu} X(t,\mu,\lambda)f(t),$$

where $X(t,\mu,\lambda)$ admits the following expansion in terms of $W$-generators:

$$(5.8) \quad X(t,\mu,\lambda) := \exp\left(\sum_1^\infty t_i(\mu^i-\lambda^i)\right) \exp\left(\sum_1^\infty (\lambda^{-i}-\mu^{-i})\frac{1}{i}\frac{\partial}{\partial t_i}\right)$$

$$= \sum_{k=0}^\infty \frac{(\mu-\lambda)^k}{k!} \sum_{\ell=-\infty}^\infty \lambda^{-\ell-k} W_\ell^{(k)}, \quad \text{with } W_\ell^{(0)} = \delta_{\ell 0}.$$

Note that $\mathbb{X}_{11}$ and $\mathbb{X}_{22}$ are closely related to the two-Toda vertex operators defined in [6]. Acting on infinite vectors of $\tau$-functions,[11] they give

$$(5.9) \quad \mathbb{X}_{11}(\mu,\lambda) = \mathbb{X}_1^* \mathbb{X}_1 = -\chi^*(\lambda)\chi(\mu)X(-t,\lambda)X(t,\mu)$$

---

[11]where $\left(\frac{\mu}{\lambda}\right)^\alpha = \sum_{k\geq 0} \binom{\alpha}{k} \left(\frac{\mu-\lambda}{\lambda}\right)^k$ and $\binom{\alpha}{k} = \frac{(\alpha)_k}{k!}$

$$= \frac{\lambda}{\mu - \lambda} \left( \left(\frac{\mu}{\lambda}\right)^n X(t, \mu, \lambda) \right)_{n \in \mathbb{Z}}$$

$$= \frac{\lambda}{\mu - \lambda} \left( \sum_{k=0}^{\infty} \frac{(\mu - \lambda)^k}{k!} \sum_{\ell=-\infty}^{\infty} \lambda^{-\ell-k} W_{n,\ell}^{(k)} \right)_{n \in \mathbb{Z}}$$

$$= \frac{\lambda}{\mu - \lambda} \mathbb{X}(t, \lambda, \mu),$$

(5.10)
$$\mathbb{X}_{22}(\mu, \lambda) = \mathbb{X}_2^* \mathbb{X}_2 = -\Lambda^{-1} \chi(\lambda) \chi^*(\mu) X(-s, \lambda) X(s, \mu) \Lambda$$

$$= \frac{\mu}{\mu - \lambda} \left( \left(\frac{\lambda}{\mu}\right)^n X(s, \mu, \lambda) \right)_{n \in \mathbb{Z}}$$

$$= \frac{\mu}{\mu - \lambda} \left( \sum_{k=0}^{\infty} \frac{(\mu - \lambda)^k}{k!} \sum_{\ell=-\infty}^{\infty} \lambda^{-\ell-k} \tilde{W}_{n,\ell}^{(k)} \right)_{n \in \mathbb{Z}}$$

$$= \frac{\mu}{\mu - \lambda} \tilde{\mathbb{X}}(s, \lambda, \mu)$$

with[12]

$$W_{n,\ell}^{(k)} = \sum_{j=0}^{k} \binom{n}{j} (k)_j W_\ell^{(k-j)} \quad \text{and} \quad \tilde{W}_{n,\ell}^{(k)} = W_{-n,\ell}^{(k)} \bigg|_{t \to s}.$$

*Remark* 2. One easily computes from (5.8) and (5.9):

$$W_n^{(0)} = \delta_{n,0}, \quad W_n^{(1)} = J_n^{(1)} \quad \text{and} \quad W_n^{(2)} = J_n^{(2)} - (n+1)J_n^{(1)}, \qquad n \in \mathbb{Z}$$

$$J_n^{(1)} := \begin{cases} \partial/\partial t_n & \text{if } n > 0 \\ (-n)t_{-n} & \text{if } n < 0 \\ 0 & \text{if } n = 0 \end{cases}, \quad J_n^{(2)} := \sum_{i+j=n} :J_i^{(1)} J_j^{(1)}:$$

and
(5.11)
$$W_{m,i}^{(1)} = W_i^{(1)} + mW_i^{(0)} \qquad W_{m,i}^{(2)} = W_i^{(2)} + 2mW_i^{(1)} + m(m-1)W_i^{(0)}$$
$$= J_i^{(1)} + m\delta_{i0}, \qquad \qquad = J_i^{(2)} + (2m - i - 1)J_i^{(1)} + m(m-1)\delta_{i0}.$$

Before establishing Theorem 5.1 we first prove the following lemmas:

---

[12]Note that in the notation of [6]

$$\mathbb{X}(t, \lambda, \mu) = \frac{\mu - \lambda}{\lambda} \mathbb{X}_{11}(\mu, \lambda) \text{ and } \tilde{\mathbb{X}}(s, \lambda, \mu) = \frac{\mu - \lambda}{\mu} \mathbb{X}_{22}(\mu, \lambda).$$

LEMMA 5.3. *The following hold*:

(i) $\quad \Psi_{1,n}^*(\lambda)\Psi_{1,n}(\mu) \;=\; -\left(\frac{\mu}{\lambda}\right)^{n+1}\frac{X(-t,\lambda)X(t,\mu)\tau_{n+1}}{\tau_{n+1}} + \left(\frac{\mu}{\lambda}\right)^n \frac{X(-t,\lambda)X(t,\mu)\tau_n}{\tau_n},$

(ii) $\quad \Psi_{2,n}^*(\lambda^{-1})\Psi_{2,n}(\mu^{-1}) \;=\; \left(\frac{\lambda}{\mu}\right)^n \frac{X(s,\mu)X(-s,\lambda)\tau_n}{\tau_n} - \left(\frac{\lambda}{\mu}\right)^{n+1}\frac{X(s,\mu)X(-s,\lambda)\tau_{n+1}}{\tau_{n+1}},$

(iii) $\quad \Psi_{2,n}^*(\lambda^{-1})\Psi_{1,n}(\mu) \;=\; (\mu\lambda)^n \frac{X(-s,\lambda)X(t,\mu)\tau_n}{\tau_{n+1}} - (\mu\lambda)^{n-1}\frac{X(-s,\lambda)X(t,\mu)\tau_{n-1}}{\tau_n},$

(iv) $\quad \Psi_{1,n-1}^*(\lambda)\Psi_{2,n-1}(\mu^{-1}) \;=\; (\mu\lambda)^{-n+1}\frac{X(-t,\lambda)X(s,\mu)\tau_n}{\tau_{n-1}} - (\mu\lambda)^{-n}\frac{X(-t,\lambda)X(s,\mu)\tau_{n+1}}{\tau_n}.$

*Proof.* (i) By the explicit expression for $X(t,\lambda)$, the right-hand side of the first relation equals

$$-\left(\frac{\mu}{\lambda}\right)^n \frac{1}{1-\frac{\mu}{\lambda}} e^{\sum_1^\infty t_i(\mu^i - \lambda^i)} \left( \frac{\mu}{\lambda} \frac{\tau_{n+1}(t + [\lambda^{-1}] - [\mu^{-1}], s)}{\tau_{n+1}(t,s)} - \frac{\tau_n(t + [\lambda^{-1}] - [\mu^{-1}], s)}{\tau_n(t,s)} \right)$$

$$= \left(\frac{\mu}{\lambda}\right)^n e^{\sum_1^\infty t_i(\mu^i - \lambda^i)} \frac{\tau_{n+1}(t + [\lambda^{-1}], s)\tau_n(t - [\mu^{-1}], s)}{\tau_{n+1}(t,s)\tau_n(t,s)}$$

$$= \Psi_{1,n}^*(\lambda)\Psi_{1,n}(\mu)$$

by Corollary 3.7 with $v_2 \to 0$, $v_1 \to 0$, $z_1 = \lambda$ and $z_2 = \mu$, and finally by (2.10).

(ii) Similarly

$$\left(\frac{\lambda}{\mu}\right)^n e^{\sum_1^\infty s_i(\mu^i - \lambda^i)} \frac{1}{1-\frac{\lambda}{\mu}} \left( -\frac{\lambda}{\mu} \frac{\tau_{n+1}(t, s + [\lambda^{-1}] - [\mu^{-1}])}{\tau_{n+1}(t,s)} + \frac{\tau_n(t, s + [\lambda^{-1}] - [\mu^{-1}])}{\tau_n(t,s)} \right)$$

$$= \left(\frac{\lambda}{\mu}\right)^n e^{\sum_1^\infty s_i(\mu^i - \lambda^i)} \frac{\tau_n(t, s + [\lambda^{-1}])\tau_{n+1}(t, s - [\mu^{-1}])}{\tau_{n+1}(t,s)\tau_n(t,s)}$$

$$= \Psi_{2,n}^*(\lambda^{-1})\Psi_{2,n}(\mu^{-1})$$

by Corollary 3.7, with $z_1 \to \infty$, $z_2 \to \infty$, $v_1 = \lambda^{-1}$ and $v_2 = \mu^{-1}$, and by (2.10).

(iii) Also, the right-hand side of the third relation equals

$$(\mu\lambda)^n e^{\sum_1^\infty (t_i\mu^i - s_i\lambda^i)} \left( \frac{\tau_n(t - [\mu^{-1}], s + [\lambda^{-1}])}{\tau_{n+1}(t,s)} - \frac{1}{\mu\lambda}\frac{\tau_{n-1}(t - [\mu^{-1}], s + [\lambda^{-1}])}{\tau_n(t,s)} \right)$$

$$= (\mu\lambda)^n e^{\sum_1^\infty (t_i\mu^i - s_i\lambda^i)} \frac{\tau_n(t, s + [\lambda^{-1}])\tau_n(t - [\mu^{-1}], s)}{\tau_{n+1}(t,s)\tau_n(t,s)}$$

$$= \Psi_{2,n}^*(\lambda^{-1})\Psi_{1,n}(\mu)$$

by Corollary 3.6 with $u = 0$, $v = \lambda^{-1}$ and $z = \mu$, and by (2.10).

(iv) Finally,

$$(\mu\lambda)^{-n+1} e^{\sum_1^\infty (s_i\mu^i - t_i\lambda^i)} \left( -\frac{1}{\mu\lambda} \frac{\tau_{n+1}(t+[\lambda^{-1}], s-[\mu^{-1}])}{\tau_n(t,s)} + \frac{\tau_n(t+[\lambda^{-1}], s-[\mu^{-1}])}{\tau_{n-1}(t,s)} \right)$$

$$= (\mu\lambda)^{-n+1} e^{\sum_1^\infty (s_i\mu^i - t_i\lambda^i)} \frac{\tau_n(t+[\lambda^{-1}], s)\tau_n(t, s-[\mu^{-1}])}{\tau_n(t,s)\tau_{n-1}(t,s)}$$

$$= \Psi^*_{1,n-1}(\lambda)\Psi_{2,n-1}(\mu^{-1})$$

by Corollary 3.6 with $t \mapsto t+[\lambda^{-1}]$, $z = \lambda$, $v = 0$ and $u = \mu^{-1}$ and by (2.10). □

In the next lemma we show that the Christoffel-Darboux type kernels, formed by means of the two-Toda wave function (2.10) can be expressed in terms of vertex operator acting on the $\tau$-functions; set $\mathbb{X} := \mathbb{X}(\mu, \lambda)$:

LEMMA 5.4. *When $1 < |\mu\lambda| < |\mu|^2$, the following holds (bi-infinite case)*

(i) $\quad \sum_{j<n} \Psi^*_{1,j}(\lambda)\Psi_{1,j}(\mu) \;=\; -\left(\frac{\mu}{\lambda}\right)^n \frac{X(-t,\lambda)X(t,\mu)\tau_n}{\tau_n} = (\tau^{-1}\mathbb{X}_{11}(\tau))_n,$

(ii) $\quad \sum_{j\geq n} \Psi^*_{2,j}(\lambda^{-1})\Psi_{2,j}(\mu^{-1}) \;=\; \left(\frac{\lambda}{\mu}\right)^n \frac{X(s,\mu)X(-s,\lambda)\tau_n}{\tau_n} = (\tau^{-1}\mathbb{X}_{22}(\tau))_n,$

(iii) $\quad \sum_{j<n} \Psi^*_{2,j}(\lambda^{-1})\Psi_{1,j}(\mu) \;=\; (\mu\lambda)^{n-1} \frac{X(-s,\lambda)X(t,\mu)\tau_{n-1}}{\tau_n} = (\tau^{-1}\mathbb{X}_{12}(\tau))_n,$

(iv) $\quad \sum_{j\geq n} \Psi^*_{1,j}(\lambda)\Psi_{2,j}(\mu^{-1}) \;=\; (\mu\lambda)^{-n} \frac{X(-t,\lambda)X(s,\mu)\tau_{n+1}}{\tau_n} = (\tau^{-1}\mathbb{X}_{21}(\tau))_n.$

The *proof* is based on summing up the expressions in Lemma 5.3 and noting that, given the inequalities above,

$$\left(\frac{\mu}{\lambda}\right)^n \text{ and } (\mu\lambda)^n \to 0 \quad \text{when } n \to -\infty$$

and

$$\left(\frac{\lambda}{\mu}\right)^n \text{ and } (\mu\lambda)^{-n} \to 0 \quad \text{when } n \to +\infty.$$

*Proof of Theorem* 5.1 *and Corollary* 5.2. In the bi-infinite case, the statement of Lemma 5.4 leads at once to (5.3), whereas in the semi-infinite case, summing up the expression (i) of Lemma 5.3 yields a boundary term, by the fact that $\tau_0 = 1$. For (iii) the boundary term vanishes, since $\tau_{-1} = 0$. To prove the first expansion of the corollary, let $n \uparrow \infty$ in the (1,1)-entry of (5.4), assuming $|\mu/\lambda| < 1$. Setting $n = 0$ in the (2,2) and (1,2)-entries of (5.4) yields the second and third relations of (5.5), after first stripping the exponential part from the equation. □

## 6. A remarkable trace formula

Given two differential polynomials $p(\partial_t)$ and $q(\partial_s)$, define the customary Hirota operation:

$$(6.1) \quad p(\partial_t)q(\partial_s)f \circ g := p\left(\frac{\partial}{\partial y}\right)q\left(\frac{\partial}{\partial z}\right)f(t+y,s+z)g(t-y,s-z)\Big|_{y=z=0}.$$

THEOREM 6.1. *In the semi-infinite case, we have the following trace formula involving elementary Schur polynomials, for $n, m \geq -1$:*

$$(6.2) \quad \sum_{0 \leq i \leq N-1} (L_1^{n+1} L_2^{m+1})_{ii} = \frac{1}{\tau_N(t,s)} p_{n+N}(\tilde{\partial}_t) p_{m+N}(-\tilde{\partial}_s)\tau_1 \circ \tau_{N-1}.$$

*Remark* 1. Here is an alternative way of writing (6.2):

$$(6.3) \quad \sum_{0 \leq i \leq N-1} (L_1^{n+1} L_2^{m+1})_{ii} = \frac{1}{\tau_N(t,s)} \sum_{\substack{i+i'=n+N \\ j+j'=m+N \\ i,i',j,j' \geq 0}} \mu_{ij} p_{i'}(\tilde{\partial}_t) p_{j'}(-\tilde{\partial}_s)\tau_{N-1}(t,s)$$

where the $\mu_{ij}$ are the moments

$$\mu_{ij} := \iint x^i y^j e^{V(x,y)} dx\, dy.$$

We shall need the following matrix of matrix operators:[13]

$$(6.4)$$
$$\begin{pmatrix} N_{11} & N_{12} \\ N_{21} & N_{22} \end{pmatrix} = \begin{pmatrix} W_1 e^{(\mu-\lambda)\varepsilon}\delta(\Lambda/\lambda)W_1^{-1} & W_1 e^{(\mu^{-1}-\lambda)\varepsilon^*}\delta(\Lambda^*/\lambda)W_2^{-1} \\ -W_2 e^{(\mu^{-1}-\lambda)\varepsilon}\delta(\Lambda/\lambda)W_1^{-1} & -W_2 e^{(\mu-\lambda)\varepsilon^*}\delta(\Lambda^*/\lambda)W_2^{-1} \end{pmatrix}$$

where $\delta(\Lambda/\lambda) = \lambda\delta(\lambda,\Lambda)$, with the delta function $\delta(z) = \sum_{n:=-\infty}^{\infty} z^n$, as defined in (1.3).

PROPOSITION 6.2. *The following holds*:

$$(6.5)$$
$$\begin{pmatrix} N_{11} & N_{12} \\ N_{21} & N_{22} \end{pmatrix} := \begin{pmatrix} \Psi_1(\mu) \otimes \Psi_1^*(\lambda) & \Psi_1(\mu) \otimes \Psi_2^*(\lambda^{-1}) \\ -\Psi_2(\mu^{-1}) \otimes \Psi_1^*(\lambda) & -\Psi_2(\mu^{-1}) \otimes \Psi_2^*(\lambda^{-1}) \end{pmatrix}$$
$$= \left(\Psi_1(\mu), -\Psi_2(\mu^{-1})\right) \otimes \left(\Psi_1^*(\lambda), \Psi_2^*(\lambda^{-1})\right).$$

*Proof.* Using (1.4) and Lemma 1.1, we compute
$$\begin{aligned} N_{11} &= \lambda W_1 e^{(\mu-\lambda)\varepsilon}\delta(\lambda,\Lambda)W_1^{-1} \\ &= W_1 \chi(\mu) \otimes \chi^*(\lambda) W_1^{-1} \\ &= (W_1\chi(\mu)) \otimes ((W_1^\top)^{-1}\chi^*(\lambda)) = \Psi_1(\mu) \otimes \Psi_1^*(\lambda) \end{aligned}$$

---

[13] Note that in the notation of [6], $N_1 = \frac{\mu-\lambda}{\lambda} N_{11}$ and $N_2 = -\frac{\mu-\lambda}{\lambda} N_{22}$.

and

$$\begin{aligned} N_{22} &= -\lambda W_2 \, e^{(\mu-\lambda)\varepsilon^*}\delta(\lambda,\Lambda^*)\, W_2^{-1} \\ &= -W_2 \, \chi^*(\mu) \otimes \chi(\lambda)\, W_2^{-1} \\ &= -(W_2\chi(\mu^{-1})) \otimes ((W_2^\top)^{-1}\chi^*(\lambda^{-1})) = -\Psi_2(\mu^{-1}) \otimes \Psi_2^*(\lambda^{-1}). \end{aligned}$$

The remaining relations are established in a similar way. □

*Remark.* The operators $N(t,\mu,\lambda)$ have been considered in [6]; they are generating functions of symmetries on the $\Psi$-manifold in the following sense, by (2.8):

$$\begin{aligned} N_{ii} &= (-1)^{i-1}\lambda e^{(\mu-\lambda)M_i}\delta(\lambda, L_i) \\ &= (-1)^{i-1}\frac{\lambda}{\mu-\lambda}\sum_{k=1}^{\infty}\frac{(\mu-\lambda)^k}{k!}\sum_{\ell=-\infty}^{\infty}\lambda^{-\ell-k}k(M_i^{k-1}L_i^{k-1+\ell}). \end{aligned}$$

PROPOSITION 6.3. *In the bi-infinite case, the following holds*:

(6.6)
$$\begin{aligned} \sum_{n\in\mathbb{Z}} z^{-n}L_1^n &= \Psi_1(z) \otimes \Psi_1^*(z), \\ \sum_{n\in\mathbb{Z}} z^{-n}L_2^n &= \Psi_2(z^{-1}) \otimes \Psi_2^*(z^{-1}). \end{aligned}$$

*Proof.* Setting $\mu=\lambda=z$ in formula (6.4) for $N_{ii}$, we find

$$N_{ii}(t,z,z) = (-1)^{i-1}\delta(L_i/z) = (-1)^{i-1}\sum_{n\in\mathbb{Z}} z^{-n}L_i^n,$$

which combined with Proposition 6.2 yields Proposition 6.3. □

PROPOSITION 6.4. *In the semi-infinite case, the following holds (see notation* (2.14)):

(6.7)
$$\begin{aligned} \sum_{n\in\mathbb{Z}} z^{-n}L_1^{(n)} &= \Psi_1(z) \otimes \Psi_1^*(z), \\ \sum_{n\in\mathbb{Z}} z^{-n}L_2^{\top(n)} &= \Psi_2^*(z^{-1}) \otimes \Psi_2(z^{-1}). \end{aligned}$$

*Proof.* Acting with $\sum_{n\in\mathbb{Z}} z_1^{-n}L_1^{(n)}$ on the wave function $\Psi_1(z)$, recalling that $\pi_+$ is the projection $\pi_+(\sum_{i\in\mathbb{Z}}) = \sum_{i\geq 0} a_i z^i$, and using the usual property of the $\delta$-function in the third equality, we obtain:

$$\begin{aligned} \left(\sum_{n\in\mathbb{Z}} z_1^{-n}L_1^{(n)}\right)\Psi_1(z)e^{-\sum_1^\infty t_i z^i} &= \sum_{n\in\mathbb{Z}} z_1^{-n}\pi_+\left(z^n\Psi_1(z)e^{-\sum_1^\infty t_i z^i}\right) \\ &= \pi_+\left(\sum_{n\in\mathbb{Z}}\left(\frac{z}{z_1}\right)^n \Psi_1(z)e^{-\sum_1^\infty t_i z^i}\right) \end{aligned}$$

$$
\begin{aligned}
&= \pi_+ \left( \sum_{n\in\mathbb{Z}} \left(\frac{z}{z_1}\right)^n \Psi_1(z_1) e^{-\sum_1^\infty t_i z_1^i} \right) \\
&= \Psi_1(z_1) e^{-\sum_1^\infty t_i z_1^i} \pi_+ \left( \sum_{n\in\mathbb{Z}} \left(\frac{z}{z_1}\right)^n \right) \\
&= \Psi_1(z_1) e^{-\sum_1^\infty t_i z_1^i} \frac{1}{1-\frac{z}{z_1}} \\
&= \Psi_1(z_1) e^{-\sum_1^\infty t_i z^i} \frac{e^{\sum_1^\infty t_i(z^i - z_1^i)}}{1-\frac{z}{z_1}} \\
&= \Psi_1(z_1) \sum_{j=0}^{\infty} \Psi_{1,j}^*(z_1) \Psi_{1,j}(z) e^{-\sum_1^\infty t_i z^i} \\
&\qquad\qquad\qquad\qquad\qquad \text{by (5.5)} \\
&= (\Psi_1(z_1) \otimes \Psi_1^*(z_1)) \Psi_1(z) e^{-\sum_1^\infty t_i z^i},
\end{aligned}
$$

which is valid for all $z, z_1 \in \mathbb{C}$; this establishes the first relation of Proposition 6.4.

Similarly, one shows for all $z, z_2 \in \mathbb{C}$

$$
\begin{aligned}
\left(\sum_{n\in\mathbb{Z}} z_2^{-n} L_2^{\top(n)}\right) \Psi_2^*(z^{-1}) e^{\sum_1^\infty s_i z^i} &= \Psi_2^*(z_2^{-1}) e^{\sum_1^\infty s_i z_2^i} \frac{1}{1-z/z_2} \\
&= \Psi_2^*(z_2^{-1}) \sum_{j=0}^{\infty} \Psi_{2j}^*(z^{-1}) \Psi_{2j}(z_2^{-1}) e^{\sum_1^\infty s_i z^i} \\
&\qquad\qquad\qquad\qquad\qquad \text{by (5.5)} \\
&= (\Psi_2^*(z_2^{-1}) \otimes \Psi_2(z_2^{-1})) \Psi_2^*(z^{-1}) e^{\sum_1^\infty s_i z^i},
\end{aligned}
$$

leading to the second relation. $\square$

*Proof of Theorem* 6.1. Multiplying the first relation of Proposition 6.3 with the second transposed, and using Proposition 6.4 in the second equality, we find

$$
\begin{aligned}
\sum_{n,m\in\mathbb{Z}} \lambda^{-n} \mu^{-m} L_1^{(n)} (L_2^{\top(m)})^\top &= \sum_{n\in\mathbb{Z}} \lambda^{-n} L_1^{(n)} \sum_{m\in\mathbb{Z}} \mu^{-m} (L_2^{\top(m)})^\top \\
&= (\Psi_1(\lambda) \otimes \Psi_1^*(\lambda))(\Psi_2(\mu^{-1}) \otimes \Psi_2^*(\mu^{-1})) \\
&= \Psi_1(\lambda) \otimes \Psi_2^*(\mu^{-1}) \langle \Psi_1^*(\lambda), \Psi_2(\mu^{-1})\rangle,
\end{aligned}
$$

using regular matrix multiplication, in the last equality.

Upon taking the trace of the matrix above up to $N - 1$, and using both Theorem 5.1 and Corollary 5.2, we find

$$(6.8) \quad \tau_N(t,s) \sum_{n,m \in \mathbb{Z}} \lambda^{-n} \mu^{-m} \sum_{0 \leq i \leq N-1} (L_1^{(n)} (L_2^{\top(m)})^\top)_{ii}$$

$$= \tau_N(t,s) \sum_{0 \leq i \leq N-1} \Psi_{1i}(\lambda) \Psi_{2i}^*(\mu^{-1}) \cdot \langle \Psi_1^*(\lambda), \Psi_2(\mu^{-1}) \rangle$$

$$= (\lambda \mu)^{N-1} e^{\sum_0^\infty (t_n \lambda^n - s_n \mu^n)} \tau_{N-1}(t - [\lambda^{-1}], s + [\mu^{-1}])$$

$$\cdot e^{\sum_0^\infty (s_n \mu^n - t_n \lambda^n)} \tau_1(t + [\lambda^{-1}], s - [\mu^{-1}])$$

$$= (\lambda \mu)^{N-1} \tau_1(t + [\lambda^{-1}], s - [\mu^{-1}]) \tau_{N-1}(t - [\lambda^{-1}], s + [\mu^{-1}]).$$

Then, using the Taylor expansion, in $\lambda^{-1}$ and $\mu^{-1}$, we find:

$$f(t \pm [\lambda^{-1}], s \mp [\mu^{-1}]) = \sum_{n=0}^\infty p_n(\pm \tilde{\partial}_t) f(t, s \mp [\mu^{-1}]) \lambda^{-n}$$

$$= \sum_{n=0}^\infty p_n(\pm \tilde{\partial}_t) \left( \sum_{m=0}^\infty p_m(\mp \tilde{\partial}_s) f(t,s) \mu^{-m} \right) \lambda^{-n}$$

$$= \sum_{m,n=0}^\infty (p_n(\pm \tilde{\partial}_t) p_m(\mp \tilde{\partial}_s) f(t,s)) \lambda^{-n} \mu^{-m}.$$

Therefore, on the one hand,

$$(6.9) \quad \tau_1(t + [\lambda^{-1}], s - [\mu^{-1}]) = \sum_{m,n=0}^\infty p_n(\tilde{\partial}_t) p_m(-\tilde{\partial}_s) \tau_1(t,s) \lambda^{-n} \mu^{-m};$$

on the other hand, using the explicit matrix representation of $\tau_1$ (see [2]), we see that

(6.10)

$$\tau_1(t+[\lambda^{-1}], s-[\mu^{-1}]) = \iint e^{\sum_1^\infty (t_i + \frac{1}{i\lambda^i})x^i - (s_i - \frac{1}{i\mu^i})y^i} e^{V_0(x,y)} dx\, dy$$

$$= \iint \frac{e^{\sum_1^\infty (t_i x^i - s_i y^i)}}{(1 - \frac{x}{\lambda})(1 - \frac{y}{\mu})} e^{V_0(x,y)} dx\, dy$$

$$= \sum_{m,n \geq 0} \lambda^{-n} \mu^{-m} \iint x^n y^m e^{V_{t,s}(x,y)} dx\, dy$$

$$= \sum_{m,n \geq 0} \mu_{n,m} \lambda^{-n} \mu^{-m},$$

which, upon comparison with (6.9), leads to

$$\mu_{n,m}(t,s) = p_n(\tilde{\partial}_t) p_m(-\tilde{\partial}_s) \tau_1(t,s).$$

Therefore (6.8) reads

$$\tau_N(t,s) \sum_{n,m\in\mathbb{Z}} \lambda^{-n}\mu^{-m} \sum_{0\le i\le N-1} (L_1^{(n)}(L_2^{\top(m)})^\top)_{ii}$$

$$= (\lambda\mu)^{N-1} \sum_{i,j\ge 0} p_i(\tilde{\partial}_t)p_j(-\tilde{\partial}_s)\tau_1(t,s)\lambda^{-i}\mu^{-j}$$

$$\cdot \sum_{i',j'\ge 0} p_{i'}(-\tilde{\partial}_t)p_{j'}(\tilde{\partial}_s)\tau_{N-1}(t,s)\lambda^{-i'}\mu^{-j'}$$

$$= \sum_{n,m\in\mathbb{Z}} \lambda^{-n}\mu^{-m} \sum_{\substack{i+i'=N+n-1 \\ j+j'=N+m-1}} p_i(\tilde{\partial}_t)p_j(-\tilde{\partial}_s)\tau_1(t,s)p_{i'}(-\tilde{\partial}_t)p_{j'}(\tilde{\partial}_s)\tau_{N-1}(t,s).$$

Upon comparison of the coefficients of $\lambda^{-n}\mu^{-m}$ for $n,m\ge 0$, we find

$$\tau_N(t,s)\sum_{0\le i\le N-1}(L_1^n L_2^m)_{ii}$$

$$= \tau_N(t,s)\sum_{0\le i\le N-1}(L_1^{(n)}(L_2^{\top(m)})^\top)_{ii},$$

$$\text{since } L_1^{(n)} = L_1^n, L_2^{\top(m)} = (L_2^\top)^m \text{ for } n,m\ge 0,$$

$$= \sum_{\substack{i+i'=N+n-1 \\ j+j'=N+m-1}} p_i(\tilde{\partial}_t)p_j(-\tilde{\partial}_s)\tau_1(t,s)p_{i'}(-\tilde{\partial}_t)p_{j'}(\tilde{\partial}_s)\tau_{N-1}(t,s)$$

$$= p_{N+n-1}(\tilde{\partial}_t)p_{N+m-1}(-\tilde{\partial}_s)\tau_1\circ\tau_{N-1}, \text{ for } n,m\ge 0,$$

leading to the statement of Theorem 6.1. □

## 7. Two-Toda symmetries and the ASV-correspondence

Define the four vector fields

$$(\Psi_1,\Psi_2)^{\cdot} = \mathbb{Y}_{ij}(\Psi_1,\Psi_2) := (-N_{ij,\ell}\Psi_1, N_{ij,u}\Psi_2)$$

on the manifold of wave vectors $\Psi = (\Psi_1,\Psi_2)$ and the four vector fields

$$\dot\tau := \mathbb{X}_{ij}\tau$$

on the manifold of $\tau$-vectors. They are symmetries of the two-Toda lattice; i.e., they commute with the basic $(t,s)$-flows (see [6]).

THEOREM 7.1. *There exists the Adler-Shiota-van Moerbeke-correspondence between symmetry vector fields on $\Psi$ and those acting on $\tau$:*

(7.1) $$\left(-\frac{N_{ij,\ell}\Psi_1}{\Psi_1}, \frac{N_{ij,u}\Psi_2}{\Psi_2}\right) = \left((e^{-\eta_1}-1)\frac{\mathbb{X}_{ij}\tau}{\tau}, (\Lambda e^{-\eta_2}-1)\frac{\mathbb{X}_{ij}\tau}{\tau}\right).$$

*Proof.* For $N_{11}$ and $N_{22}$, the result follows from [6] by the conversion rule from $N_i$ to $N_{ii}$ (see footnote 13) and from $(\mathbb{X}(t,\lambda,\mu), \mathbb{X}(s,\lambda,\mu))$ to $(\mathbb{X}_{11}, \mathbb{X}_{22})$ (see footnote 12). For $i\ne j$, we prove for instance the following two identities:

(i) $-\dfrac{N_{12\ell}\Psi_1(z)}{\Psi_1(z)} = (e^{-\eta_1} - 1)\dfrac{\mathbb{X}_{12}\tau}{\tau}.$

For the vector fields $\dot{\tau}_n = \mathbb{X}_{12}\tau_n$ acting on the $\tau$-functions $\tau_n$, consider the derivation of $\Psi_1$ with regard to that vector field:

$$(7.2) \qquad \dfrac{(\dot{\Psi}_1(z))_n}{\Psi_n(z)} = \left(\dfrac{e^{-\eta_1}\tau_n(t,s)}{\tau_n(t,s)}\right)^{\cdot} \dfrac{\tau_n(t,s)}{e^{-\eta_1}\tau_n(t,s)}$$

$$= \dfrac{(e^{-\eta_1}\dot{\tau}_n(t,s))\tau_n(t,s) - e^{-\eta_1}\tau_n(t,s)\dot{\tau}_n(t,s)}{e^{-\eta_1}\tau_n(t,s)\tau_n(t,s)}$$

$$= (e^{-\eta_1} - 1)\dfrac{\mathbb{X}_{12}\tau_n(t,s)}{\tau_n(t,s)},$$

where in the above we have used the commutation of the symmetries with the $t$-flows. Considering the vector field $\dot{f} = \mathbb{X}_{12}f$ (see (5.2)) acting on column vectors $f = (f_n(t,s))_{n \in \mathbb{Z}}$, we compute using (7.2) and, in the third identity, the relation of Corollary 3.7, with $n \mapsto n-1$, $t \mapsto t - [z^{-1}]$, $s \mapsto s + [\lambda^{-1}]$, $z_1 = z$, $z_2 = \mu$, $v_1 = 0$, $v_2 = \lambda^{-1}$,

$$\dfrac{(\dot{\Psi}_1(z))_n}{\Psi_{1n}(\mu)} = \dfrac{e^{\Sigma t_i z^i} z^n \Big((\mathbb{X}_{12}\tau(t-[z^{-1}],s))_n \tau_n(t,s) - (\mathbb{X}_{12}\tau(t,s))_n \tau_n(t-[z^{-1}],s)\Big)}{e^{\Sigma t_i \mu^i} \mu^n \tau_n(t,s)\tau_n(t-[\mu^{-1}],s)}$$

$$= e^{\Sigma t_i(z^i - \mu^i)} \mu e^{\Sigma_1^\infty (t_i \mu^i - s_i \lambda^i)} \left(\dfrac{z}{\mu}\right)^n (\mu\lambda)^{n-1}$$

$$\times \dfrac{(\mu^{-1} - z^{-1})\tau_{n-1}(t-[z^{-1}]-[\mu^{-1}],s+[\lambda^{-1}])\tau_n(t,s) - \mu^{-1}\tau_{n-1}(t-[\mu^{-1}],s+[\lambda^{-1}])\tau_n(t-[z^{-1}],s)}{\tau_n(t,s)\tau_n(t-[\mu^{-1}],s)}$$

$$= -e^{\Sigma(t_i z^i - s_i \lambda^i)} z^n \lambda^{n-1} z^{-1} \dfrac{\tau_{n-1}(t-[z^{-1}], s+[\lambda^{-1}])\tau_n(t-[\mu^{-1}],s)}{\tau_n(t,s)\tau_n(t-[\mu^{-1}],s)}$$

$$= -(\lambda z)^{n-1} \dfrac{X(-s,\lambda)X(t,z)\tau_{n-1}}{\tau_n}, \text{ using the definition of } X(t,z)$$

$$= -\sum_{j<n} \Psi_{2j}^*(\lambda^{-1})\Psi_{1j}(z), \text{ by Lemma 5.4 (iii),}$$

$$= -\dfrac{\sum_{j<n}\Psi_{1n}(\mu)\Psi_{2j}^*(\lambda^{-1})\Psi_{1j}(z)}{\Psi_{1n}(\mu)}$$

$$= -\dfrac{(N_{12\ell}\Psi_1(z))_n}{\Psi_{1n}(\mu)}, \text{ by Proposition 6.2,}$$

thus proving

$$(7.3) \qquad (\dot{\Psi}_1(z))_n = -(N_{12\ell}\Psi_1(z))_n.$$

Comparison of both expressions (7.2) and (7.3) for $\dot{\Psi}_1(z)$ yields (i).

(ii) $\dfrac{N_{21u}\Psi_2}{\Psi_2} = (\Lambda e^{-\eta_2} - 1)\dfrac{\mathbb{X}_{21}\tau}{\tau}.$

Consider now the derivative of $\Psi_2$ with regard to the vector field $\dot{\tau}_n = \mathbb{X}_{21}\tau_n$, acting on $\tau$-functions. At first

$$
\begin{aligned}
(7.4) \quad \dfrac{(\dot{\Psi}_2(z))_n}{\Psi_{2n}(z)} &= \left(\dfrac{e^{-\eta_2}\tau_{n+1}(t,s)}{\tau_n(t,s)}\right)^{\cdot} \dfrac{\tau_n(t,s)}{e^{-\eta_2}\tau_{n+1}(t,s)} \\
&= \dfrac{(e^{-\eta_2}\dot{\tau}_{n+1}(t,s))\tau_n(t,s) - (e^{-\eta_2}\tau_{n+1}(t,s))\dot{\tau}_n(t,s)}{e^{-\eta_2}\tau_{n+1}(t,s)\tau_n(t,s)} \\
&= (e^{-\eta_2}\Lambda - 1)\left(\dfrac{\mathbb{X}_{21}\tau}{\tau}\right)_n.
\end{aligned}
$$

Acting with the vector field $\dot{f} = \mathbb{X}_{21}f$ on $\Psi_2(z)$, and using (7.4), we apply, in the third identity, Corollary 3.6 with $n \mapsto n+1$, with $t \mapsto t+[z^{-1}]$, $s \mapsto s-[v]$ and subsequently with $z \mapsto \lambda$, $u \mapsto \mu^{-1}$, $v \mapsto z$:

$$
\begin{aligned}
-\dfrac{(\dot{\Psi}_2(z))_n}{\Psi_{2,n}(\mu^{-1})} &= -\dfrac{e^{\sum s_i z^{-i}} z^n \left((\mathbb{X}_{21}\tau(t,s-[z]))_{n+1}\tau_n(t,s) - (\mathbb{X}_{21}\tau(t,s))_n \tau_{n+1}(t,s-[z])\right)}{e^{\sum s_i \mu^i} \mu^{-n} \tau_n(t,s)\tau_{n+1}(t,s-[\mu^{-1}])} \\
&= -e^{\sum s_i(z^{-i}-\mu^i)} e^{\sum(s_i\mu^i - t_i\lambda^i)} (\mu z)^n \left(\dfrac{1}{\mu\lambda}\right)^{n+1} \\
&\quad \times \dfrac{(1-z\mu)\tau_{n+2}(t+[\lambda^{-1}],s-[z]-[\mu^{-1}])\tau_n(t,s) - \mu\lambda\tau_{n+1}(t+[\lambda^{-1}],s-[\mu^{-1}])\tau_{n+1}(t,s-[z])}{\tau_n(t,s)\tau_{n+1}(t,s-[\mu^{-1}])} \\
&= e^{\sum s_i z^{-i} - \sum t_i \lambda^i} \left(\dfrac{z}{\lambda}\right)^n \dfrac{\mu\lambda\tau_{n+1}(t+[\lambda^{-1}],s-[z])\tau_{n+1}(t,s-[\mu^{-1}])}{\mu\lambda\tau_n(t,s)\tau_{n+1}(t,s-[\mu^{-1}])} \\
&= (\lambda z^{-1})^{-n} \dfrac{X(-t,\lambda)X(s,z^{-1})\tau_{n+1}(t,s)}{\tau_n(t,s)} \\
&= \sum_{j \geq n} \Psi^*_{1j}(\lambda)\Psi_{2j}(z), \text{ from Lemma 5.4 (iv),} \\
&= \dfrac{\sum_{j \geq n} \Psi_{2n}(\mu^{-1})\Psi^*_{1j}(\lambda)\Psi_{2j}(z)}{\Psi_{2n}(\mu^{-1})} \\
&= -\dfrac{(N_{21u}\Psi_2(z))_n}{\Psi_{2n}(\mu^{-1})}, \text{ by Proposition 6.2,}
\end{aligned}
$$

and thus
$$(\dot{\Psi}(z))_n = (N_{21u}\Psi_2(z))_n.$$

Comparison with expression (7.4) yields (ii). The proofs of the other identities contained in (7.1) can be done in the same style. Note that even the identities involving $N_{11}$ and $N_{22}$, which were established in [6], can be shown in this fashion. This ends the proof of Theorem 7.1. □

## 8. Fredholm determinants of Christoffel-Darboux kernels

The following theorem involves determinants of the kernels:

$$K_{11,n}(y,z) := \sum_{\ell<n} \Psi^*_{1\ell}(z)\Psi_{1\ell}(y), \qquad K_{21,n}(y,z) := \sum_{\ell\geq n} \Psi^*_{1\ell}(z)\Psi_{2\ell}(y^{-1}),$$

$$K_{12,n}(y,z) := \sum_{\ell<n} \Psi^*_{2\ell}(z^{-1})\Psi_{1\ell}(y), \qquad K_{22,n}(y,z) := \sum_{\ell\geq n} \Psi^*_{2\ell}(z^{-1})\Psi_{2\ell}(y^{-1}),$$

already mentioned in Theorem 5.1:

THEOREM 8.1.  *The following holds*:

(8.1)
$$\begin{pmatrix} \left(\det(K_{11,n}(y_i,z_j))_{1\leq i,j\leq k}\right)_{n\in\mathbb{Z}} & \left(\det(K_{21,n}(y_i,z_j))_{1\leq i,j\leq k}\right)_{n\in\mathbb{Z}} \\ \left(\det(K_{12,n}(y_i,z_j))_{1\leq i,j\leq k}\right)_{n\in\mathbb{Z}} & \left(\det(K_{22,n}(y_i,z_j))_{1\leq i,j\leq k}\right)_{n\in\mathbb{Z}} \end{pmatrix}$$
$$= \frac{1}{\tau}\begin{pmatrix} \prod_{\ell=1}^k \mathbb{X}_{11}(y_\ell,z_\ell)\tau & \prod_{\ell=1}^k \mathbb{X}_{21}(y_\ell,z_\ell)\tau \\ \prod_{\ell=1}^k \mathbb{X}_{12}(y_\ell,z_\ell)\tau & \prod_{\ell=1}^k \mathbb{X}_{22}(y_\ell,z_\ell)\tau \end{pmatrix}.$$

*Proof.* We work out the result for the kernels $K_{12}$ and $K_{11}$. Indeed, since, using Lemma 5.4,

$$K_{12,n}(z_j,y_i) = \sum_{\ell<n} \Psi^*_{2\ell}(y_i^{-1})\Psi_{1\ell}(z_j)$$
$$= (y_i z_j)^{n-1} e^{\sum_{\alpha=1}^\infty (t_\alpha z_j^\alpha - s_\alpha y_i^\alpha)} \frac{\tau_{n-1}(t-[z_j^{-1}], s+[y_i^{-1}])}{\tau_n(t,s)}$$

we have, using the second relation (4.1) of Theorem 4.2,

(8.2)
$$\det\left(\sum_{\ell<n} \Psi^*_{2\ell}(y_i^{-1})\Psi_{1\ell}(z_j)\right)_{1\leq i,j\leq k}$$
$$= \left(\prod_1^k y_i z_i\right)^{n-1} \left(\prod_{i=1}^k e^{\sum_{\alpha=1}^\infty (t_\alpha z_i^\alpha - s_\alpha y_i^\alpha)}\right) \det\left(\frac{\tau_{n-1}(t-[z_j^{-1}], s+[y_i^{-1}])}{\tau_n(t,s)}\right)_{1\leq i,j\leq k}$$
$$= \left(\prod_1^k y_i z_i\right)^{n-1} \Delta(y^{-1})\Delta(z^{-1})\left(\prod_{i=1}^k e^{\sum_{\alpha=1}^\infty (t_\alpha z_i^\alpha - s_\alpha y_i^\alpha)}\right)$$
$$\cdot \frac{\tau_{n-k}(t-\sum_1^k [z_i^{-1}], s+\sum_1^k [y_i^{-1}])}{\tau_n(t,s)}$$
$$\stackrel{*}{=} \frac{1}{\tau_n}\left(\prod_{i=1}^k \mathbb{X}_{12}(z_i,y_i)\tau\right)_n,$$

using the computation below. Observe indeed that compounding two operators $\mathbb{X}_{12}(z_i, y_i)$

$$\begin{aligned}(\mathbb{X}_{12}(z_1, y_1)\tau)_n &= (\Lambda^{-1}\chi(y_1)\chi(z_1)X(t, z_1)X(-s, y_1)\tau)_n \\ &= y_1^{n-1}z_1^{n-1}e^{\sum_1^\infty (t_i z_1^i - s_i y_1^i)}\tau_{n-1}(t - [z_1^{-1}], s + [y_1^{-1}])\end{aligned}$$

yields

$$\begin{aligned}&(\mathbb{X}_{12}(z_2, y_2)\mathbb{X}_{12}(z_1, y_1)\tau)_n \\ &= \left(\mathbb{X}_{12}(z_2, y_2)((y_1 z_1)^{n-1}e^{\sum_1^\infty(t_i z_1^i - s_i y_1^i)}\tau_{n-1}(t - [z_1^{-1}], s + [y_1^{-1}]))_{n\in\mathbb{Z}}\right)_n \\ &= (y_2 z_2)^{n-1}(y_1 z_1)^{n-2}\prod_{j=1}^2 e^{\sum_{i=1}^\infty(t_i z_j^i - s_i y_j^i)}(1 - \frac{z_1}{z_2})(1 - \frac{y_1}{y_2}) \\ &\quad \cdot \tau_{n-2}\left(t - \sum_1^2 [z_i^{-1}], s + \sum_1^2 [y_i^{-1}]\right) \\ &= \left(\prod_1^2 y_i z_i\right)^{n-1}\prod_{j=1}^2 e^{\sum_{i=1}^\infty(t_i z_j^i - s_i y_j^i)}\prod_{1\leq i < j \leq 2}(z_i^{-1} - z_j^{-1})\prod_{1\leq i < j \leq 2}(y_i^{-1} - y_j^{-1}) \\ &\quad \cdot \tau_{n-2}\left(t - \sum_1^2 [z_i^{-1}], s + \sum_1^2 [y_i^{-1}]\right),\end{aligned}$$

and so on; this establishes the equality ($\stackrel{*}{=}$) in (8.2) and the $K_{12}$-identity in (8.1).

Since, by Lemma 5.4,

$$\begin{aligned}K_{11,n}(z_j, y_i) &= \sum_{\ell < n}\Psi_{1\ell}^*(y_i)\Psi_{1\ell}(z_j) \\ &= -\left(\frac{z_j}{y_i}\right)^n \frac{1}{1 - \frac{z_j}{y_i}}e^{\sum_{\alpha=1}^\infty t_\alpha(z_j^\alpha - y_i^\alpha)}\frac{\tau_n(t - [z_j^{-1}] + [y_i^{-1}], s)}{\tau_n(t, s)},\end{aligned}$$

we have, using the first relation (4.1) of Theorem 4.2,

$$\begin{aligned}&\det\left(\sum_{\ell < n}\Psi_{1\ell}^*(y_i)\Psi_{1\ell}(z_j)\right)_{1\leq i,j\leq k} \\ &= (-1)^{\frac{k(k-1)}{2}}\frac{\Delta(y)\Delta(z)}{\prod_{1\leq i,j\leq n}(y_i - z_j)}(-1)^k \prod_{i=1}^k\left(\frac{z_i^n}{y_i^{n-1}}\right) \\ &\quad \cdot \prod_{j=1}^k e^{\Sigma t_\alpha(z_j^\alpha - y_j^\alpha)}\frac{\tau_n(t - \sum_1^k[z_j^{-1}] + \sum_1^k[y_i^{-1}], s)}{\tau_n(t, s)} \\ &= \frac{1}{\tau}\prod_1^k \mathbb{X}_{11}(z_i, y_i)\tau.\end{aligned}$$

In the last equality, we used the composition of the vertex operator

$$\mathbb{X}_{11}(z,y) = \mathbb{X}_1^*(t,y)\mathbb{X}_1(t,z) = \left(\frac{z^n}{y^{n-1}}\right)\frac{X(t,z,y)}{z-y}$$

several times, to yield

$$\prod_1^k \mathbb{X}_{11}(z_i, y_i)\tau = (-1)^{\frac{k(k-1)}{2}}\frac{\Delta(z)\Delta(y)}{\prod_{k,\ell}(z_k-y_\ell)}\left(\prod_{j=1}^k e^{\sum_{\alpha=1}^\infty (t_\alpha z_j^\alpha - s_\alpha y_j^\alpha)}\right)\left(\prod_{\alpha=1}^k \frac{z^n}{y^{n-1}}\right)$$

$$\cdot \tau\left(t + \sum_1^k [y_j^{-1}] - \sum_1^k [z_j^{-1}], s\right),$$

thus establishing the result, for the $K_{11}$ and $K_{12}$ components of (8.1); the remaining cases are more of the same. □

It is also interesting to compute the Fredholm determinant of the kernel $K = K_{12,n}$, namely

(8.3)
$$\det(I - \lambda K) := 1 + \sum_1^\infty \frac{(-\lambda)^k}{k!}\int\ldots\int_{E^k}\det(K(x_i,y_j))_{1\leq i,j\leq k}\prod_1^k(\rho(x_i,y_i)dx_i dy_i)$$

over a set of the form $E = E_1 \times E_2 \subset \mathbb{R}^2$ with regard to the measure $\rho(x,y)dx\,dy$.

COROLLARY 8.2. *The vector of Fredholm determinants (in the sense above) equals*

(8.4) $$\det(I - \lambda K^E) = \frac{1}{\tau}e^{-\lambda\iint_E dx\,dy\,\rho(x,y)\mathbb{X}(x,y)}\tau$$

*for the kernel $K^E = K_{12,n}(y,z)I_E(z)$, with $\mathbb{X}_{12}(x,y)$ being the corresponding vertex operator, given before Theorem 5.1.*

*Proof.* Putting the corresponding determinant obtained in Theorem 8.1 in the Fredholm formula (8.1), we find for a subset of the form $E = E_1 \times E_2 \subset \mathbb{R}^2$,

$$(\det(I - \lambda K^E))_{n \in \mathbb{Z}}$$
$$= 1 + \sum_1^\infty \frac{(-\lambda)^k}{k!}\int\ldots\int_{E^k}\det(K(x_i,y_j))_{1\leq i,j\leq k}\prod_1^k(\rho(x_i,y_i)dx_i dy_i)$$
$$= \sum_0^\infty \frac{(-\lambda)^k}{k!}\int\ldots\int_{E^k}\frac{1}{\tau}\left(\prod_1^k \mathbb{X}(x_i,y_i)\right)\tau\prod_1^k(\rho(x_i,y_i)dx_i dy_i)$$
$$= \frac{1}{\tau}\sum_{k=0}^\infty \frac{1}{k!}\left(-\lambda\iint_E \mathbb{X}(x,y)\rho(x,y)dxdy\right)^k \tau$$
$$= \frac{1}{\tau}e^{-\lambda\iint_E dx\,dy\,\rho(x,y)\mathbb{X}(x,y)}\tau.$$
□

## 9. Differential equations for vertex operators and a Virasoro algebra of central charge $c = -2$

Consider the vector of integrals

(9.1) $$\mathbb{U}_E := \int\int_E dx\, dy\, \rho(x,y) \mathbb{X}_{12}(x,y)$$

over the subset $E := [a,b] \times [c,d] \subset \mathbb{R}^2$, of the vertex operator $\mathbb{X}_{12}(x,y)$, defined in (5.2), integrated over the weight

$$\rho(x,y) dx\, dy := e^{V_{12}(x,y)} dx dy := e^{\sum_{i,j \geq 1} c_{ij} x^i y^j} dx\, dy.$$

Also consider the vector of operators

(9.2) $$\mathbb{V}_k := -b^{k+1}\frac{\partial}{\partial b} - a^{k+1}\frac{\partial}{\partial a} + \mathbb{J}_k^{(2)} + \sum_{i,j \geq 1} ic_{ij}\frac{\partial}{\partial c_{i+k,j}},$$

with

(9.3) $$\mathbb{J}_k^{(2)} = (J_{k,n}^{(2)})_{n \in \mathbb{Z}} = \frac{1}{2}\left(J_k^{(2)} + (2n+k+1)J_k^{(1)} + n(n+1)J_k^{(0)}\right)_{n \in \mathbb{Z}};$$

then the following theorem holds:

THEOREM 9.1. *For all $k \geq -1$ and $n \geq 1$,*

$$[\mathbb{V}_k, (\mathbb{U}_E)^n] = 0,$$

*with the vector $\mathbb{J}_k^{(2)}$ forming a Virasoro algebra of central charge $c = -2$:*

$$\left[\mathbb{J}_k^{(2)}, \mathbb{J}_\ell^{(2)}\right] = (k-\ell)\mathbb{J}_{k+\ell}^{(2)} + (-2)\left(\frac{k^3-k}{12}\right)\delta_{k,-\ell}.$$

Before proving Theorem 9.1, we first need a few lemmas. For the sake of later investigations on matrix integrals (symmetric and symplectic), we introduce operators depending on a real parameter $\alpha > 0$. So, define Heisenberg and Virasoro operators, depending on $\alpha$,

$$J_k^{(1)}(\alpha) =: \begin{cases} \frac{\partial}{\partial t_k} & k > 0 \\ \frac{1}{\alpha}(-k)t_{-k} & k < 0 \\ 0 & k = 0 \end{cases} \quad \text{and} \quad J_k^{(2)}(\alpha) := \sum_{i+j=k} :J_i^{(1)} J_j^{(1)}: ,$$

together with "vector operators" acting on vectors of functions $f = (f_n(t_1, t_2, \ldots))_{n \in \mathbb{Z}}$,

$$\begin{aligned}\mathbb{J}_k^{(1)}(\alpha) &= \left(J_{k,n}^{(1)}(\alpha)\right)_{n \in \mathbb{Z}} \\ &= \left(J_k^{(1)}(\alpha) + nJ_k^{(0)}\right)_{n \in \mathbb{Z}} \quad \text{and} \quad \mathbb{J}_k^{(0)} = nJ_k^{(0)} = n\delta_{0k},\end{aligned}$$

$$\begin{aligned}
\mathbb{J}_k^{(2)}(\alpha) &= \left(J_{k,n}^{(2)}(\alpha)\right)_{n\in\mathbb{Z}} \\
&= \frac{\alpha}{2}\sum_{i+j=k} :\mathbb{J}_i^{(1)}(\alpha)\mathbb{J}_j^{(1)}(\alpha): + \left(1-\frac{\alpha}{2}\right)\left((k+1)\mathbb{J}_k^{(1)}(\alpha) - k\mathbb{J}_k^{(0)}\right) \\
&= \left(\frac{\alpha}{2}\sum_{i+j=k} :J_i^{(1)}(\alpha)J_j^{(1)}(\alpha): + \left(n\alpha + (k+1)\left(1-\frac{\alpha}{2}\right)\right)J_k^{(1)}(\alpha)\right. \\
&\qquad\qquad\qquad\qquad\qquad\qquad \left. + \frac{n((n-1)\alpha+2)}{2}J_k^{(0)}\right)_{n\in\mathbb{Z}}.
\end{aligned}$$

Note $\mathbb{J}_k^{(2)}(\alpha)$ coincides for $\alpha = 1$ with $\mathbb{J}_k^{(2)}$ of (9.3). Given the vertex operator, containing a parameter $\alpha$ as well,

$$X_\alpha(u) = e^{\sum_1^\infty t_i u^i} e^{-\alpha \sum_1^\infty \frac{u^{-i}}{i}\frac{\partial}{\partial t_i}},$$

and the "vector vertex operator," remembering $\chi(z) = (\ldots, z^{-1}, 1, z, z^2, \ldots)^\top$,

$$\mathbb{X}_\alpha(u) = \Lambda^{-1} e^{\sum_1^\infty t_i u^i} e^{-\alpha \sum_1^\infty \frac{u^{-i}}{i}\frac{\partial}{\partial t_i}} \chi(u^\alpha),$$

we prove:

LEMMA 9.2. *The vertex operator $X_\alpha := X_\alpha(u)$ and $J_{k,n}^{(\ell)} := J_{k,n}^{(\ell)}(\alpha)$ satisfy the relations*:

$$\begin{aligned}
u^k X_\alpha &= [J_k^{(1)}, X_\alpha] + X_\alpha J_k^{(0)}, \\
\frac{\partial}{\partial u} u^{k+1}(X_\alpha u^n) &= J_{k,n+1}^{(2)} X_\alpha u^n - X_\alpha u^n J_{k,n}^{(2)}.
\end{aligned}$$

*Proof.* Setting $X := X_\alpha(u)$ and using

$$\alpha u^{-\beta} X = [\beta t_\beta, X] \quad \text{and} \quad u^\beta X = \left[\frac{\partial}{\partial t_\beta}, X\right], \quad \beta \geq 1,$$

one immediately checks the first relation; although the following relation holds for all $k \in \mathbb{Z}$, for brevity one checks it for $k \geq 1$,

$$\begin{aligned}
u^{k+1}\frac{\partial}{\partial u}X &= \alpha \sum_{\substack{i\geq 1 \\ -i+k\neq 0}} u^{-i+k} X \frac{\partial}{\partial t_i} + \sum_{\substack{i\geq 1 \\ i+k\neq 0}} it_i u^{i+k} X + \alpha X \frac{\partial}{\partial t_k} \\
&= \sum_{\substack{i\geq 1 \\ -i+k<0}} [(i-k)t_{i-k}, X]\frac{\partial}{\partial t_i} + \alpha \sum_{\substack{i\geq 1 \\ -i+k>0}} \left[\frac{\partial}{\partial t_{k-i}}, X\right]\frac{\partial}{\partial t_i} \\
&\quad + \sum_{i\geq\max(1,-k+1)} it_i\left[\frac{\partial}{\partial t_{i+k}}, X\right] + \alpha X\frac{\partial}{\partial t_k}
\end{aligned}$$

$$
\begin{aligned}
&= \sum_{i\geq 1}\left[it_i\frac{\partial}{\partial t_{i+k}}, X\right] + \frac{\alpha}{2}\sum_{\substack{i+j=k\\i,j\geq 1}}\left[\frac{\partial^2}{\partial t_i \partial t_j}, X\right]\\
&\quad - \frac{\alpha}{2}\sum_{\substack{i+j=k\\i,j\geq 1}}\frac{\partial}{\partial t_i}\left[\frac{\partial}{\partial t_j}, X\right] + \frac{\alpha}{2}\sum_{\substack{i+j=k\\i,j\geq 1}}\left[\frac{\partial}{\partial t_i}, X\right]\frac{\partial}{\partial t_j}\\
&\quad + \alpha X\frac{\partial}{\partial t_k}\\
&= \frac{\alpha}{2}[J_k^{(2)}, X] + \frac{\alpha}{2}\sum_{\substack{i+j=k\\i,j\geq 1}}\left[[\frac{\partial}{\partial t_i}, X], \frac{\partial}{\partial t_j}\right] + \alpha X J_k^{(1)}\\
&= \frac{\alpha}{2}[J_k^{(2)} - (k-1)J_k^{(1)}, X] + \alpha X J_k^{(1)},
\end{aligned}
$$

where in the last equality, one has used the identity

$$
\sum_{\substack{i+j=k\\i,j\geq 1}}\left[[\frac{\partial}{\partial t_i}, X], \frac{\partial}{\partial t_j}\right] = \left(\sum_{\substack{i+j=k\\i,j\geq 1}} u^{i+j}\right) X = (k-1)u^k X = (k-1)\left[\frac{\partial}{\partial t_k}, X\right].
$$

Finally, using the above and the first commutation relation of Lemma 9.2, one computes on the one hand,

$$
\begin{aligned}
\frac{\partial}{\partial u} u^{k+1} (X(u)u^{\alpha n}) &= (k+1)u^k X u^{\alpha n} + \alpha n\, u^k X u^{\alpha n}\\
&\quad + \frac{\alpha}{2}[J_k^{(2)} - (k-1)J_k^{(1)}, X u^{\alpha n}] + \alpha X u^{\alpha n} J_k^{(1)}\\
&= (n\alpha + k + 1)u^{n\alpha+k} X + \frac{\alpha}{2}\left[J_k^{(2)}, X u^{n\alpha}\right]\\
&\quad - \frac{\alpha}{2}(k-1)u^{n\alpha}\left(u^k X - X J_k^{(0)}\right) + \alpha X u^{n\alpha} J_k^{(1)},
\end{aligned}
$$

and, on the other hand,

$$
\begin{aligned}
J_{k,n+1}^{(2)} X u^{\alpha n} - X u^{\alpha n} J_{k,n}^{(2)} &= \frac{\alpha}{2}\left[J_k^{(2)}, X u^{n\alpha}\right]\\
&\quad + \left(n\alpha + (k+1) - \frac{\alpha}{2}(k+1)\right)u^{n\alpha}(u^k X - X J_k^{(0)})\\
&\quad + \alpha J_k^{(1)} X u^{n\alpha} + X u^{n\alpha} J_k^{(0)}(\alpha n + 1).
\end{aligned}
$$

Using again the first commutation relation of Lemma 9.2, we see that the two expressions are easily seen to coincide. $\square$

PROPOSITION 9.3. *Given the vector vertex operator,*

$$\mathbb{X}_\alpha(t,u) = \Lambda^{-1} e^{\sum_1^\infty t_i u^i} e^{-\alpha \sum_1^\infty \frac{u^{-i}}{i} \frac{\partial}{\partial t_i}} \chi(u^\alpha),$$

*define the generators* $\mathbb{J}_k^{(i)}(\alpha)$ *by means of*

$$z^k \mathbb{X}_\alpha(t,z) = \left[\mathbb{J}_k^{(1)}(\alpha), \mathbb{X}_\alpha(t,z)\right], \qquad \frac{\partial}{\partial z} z^{k+1} \mathbb{X}_\alpha(t,z) = \left[\mathbb{J}_k^{(2)}(\alpha), \mathbb{X}_\alpha(t,z)\right].$$

*The* $\mathbb{J}_k^{(2)}(\alpha)$*'s form a Virasoro algebra*

$$\left[\mathbb{J}_k^{(2)}(\alpha), \mathbb{J}_\ell^{(2)}(\alpha)\right] = (k-\ell)\mathbb{J}_{k+\ell}^{(2)}(\alpha) + c\left(\frac{k^3-k}{12}\right)\delta_{k,-\ell},$$

*with central charge*

$$c = 1 - 6\left(\left(\frac{\alpha}{2}\right)^{1/2} - \left(\frac{\alpha}{2}\right)^{-1/2}\right)^2.$$

*The* $\mathbb{J}_k^{(1)}(\alpha)$ *form a Heisenberg algebra, interacting with Virasoro, as follows:*

$$\left[\mathbb{J}_k^{(1)}(\alpha), \mathbb{J}_\ell^{(1)}(\alpha)\right] = \frac{k}{\alpha}\delta_{k,-\ell}$$

$$\left[\mathbb{J}_k^{(2)}(\alpha), \mathbb{J}_\ell^{(1)}(\alpha)\right] = -\ell\, \mathbb{J}_{k+\ell}^{(1)}(\alpha) + k(k+1)\left(\frac{1}{\alpha} - \frac{1}{2}\right)\delta_{k,-\ell}.$$

*Proof.* The proof follows from Lemma 9.2 and an explicit computation for the central charge. □

PROPOSITION 9.4. *The vertex operator, defined in (5.2),*

$$\mathbb{X}_{12}(u,v) = \Lambda^{-1} X(-s,v) X(t,u) \chi(u) \chi(v)$$

*leads to a Virasoro algebra of central charge* $c = -2$,

$$\frac{\partial}{\partial u} u^{k+1} \mathbb{X}_{12}(u,v) = [\mathbb{J}_k^{(2)}, \mathbb{X}_{12}(u,v)],$$

*with generators (in t)*

$$\mathbb{J}_k^{(2)} := \mathbb{J}_k^{(2)}(\alpha)\Big|_{\alpha=1} = \left(J_{k,n}^{(2)}\right)_{n\in\mathbb{Z}}$$
$$= \frac{1}{2}\left(J_k^{(2)} + (2n+k+1)J_k^{(1)} + n(n+1)J_k^{(0)}\right)_{n\in\mathbb{Z}}.$$

*Similarly, the involution* $u \leftrightarrow v$, $t \leftrightarrow -s$ *leads to the same Virasoro algebra in s, with same central charge.*

*Proof.* Noticing that one piece of $\mathbb{X}_{12}(t,s;u,v)$ is precisely $\mathbb{X}_\alpha(t;u)$ for $\alpha = 1$, we apply Proposition 9.3 for $\alpha = 1$:

$$\begin{aligned}
\frac{\partial}{\partial u} u^{k+1} \mathbb{X}_{12}(t,s;u,v) &= \left(\frac{\partial}{\partial u} u^{k+1} \mathbb{X}_1(t;u)\right) X(-s,v)\chi(v) \\
&= \left[\mathbb{J}_k^{(2)}(1), \mathbb{X}_1(t;u)\right] X(-s,v)\chi(v) \\
&= [\mathbb{J}_k^{(2)}, \mathbb{X}_{12}].
\end{aligned}$$

The central charge $c = -2$ is obtained by setting $\alpha = 1$ in the general formula for $c$ in Proposition 9.3. □

*Proof of Theorem* 9.1. We consider the vector of operators given by the double integral of a vertex operator,

$$(9.4) \qquad \int_a^b dx \int_c^d dy \frac{\partial}{\partial x}\left(x^{k+1}\mathbb{X}_{12}(x,y)\rho(x,y)\right),$$

thought of as acting on a column of functions $F(t,s,c)$. On the one hand, the integral (9.4) equals

$$\begin{aligned}
(9.5) \quad &= x^{k+1}\int_c^d dy\, \mathbb{X}_{12}(x,y)e^{V_{12}(x,y)}\Big|_{x=a}^{x=b} \\
&= \left(b^{k+1}\frac{\partial}{\partial b} + a^{k+1}\frac{\partial}{\partial a}\right)\int_a^b dx \int_c^d dy\, \mathbb{X}_{12}(x,y)e^{V_{12}(x,y)}.
\end{aligned}$$

On the other hand, by Proposition 9.4, it equals

$$\begin{aligned}
(9.6) \quad &= \int_c^d dy \int_a^b dx \left(\frac{\partial}{\partial x}x^{k+1}\mathbb{X}_{12}(x,y)\right)\rho(x,y) \\
&\quad + \int_c^d dy \int_a^b dx\, \mathbb{X}_{12}(x,y)x^{k+1}\left(\sum_{i,j\geq 1} ic_{ij}x^{i-1}y^j\right)\rho(x,y) \\
&= \left([\mathbb{J}_k^{(2)}, \cdot] + \sum_{i,j\geq 1} ic_{ij}\frac{\partial}{\partial c_{i+k,j}}\right)\int_c^d dy \int_a^b dx\, \mathbb{X}_{12}(x,y)e^{V_{12}(x,y)},
\end{aligned}$$

where the derivations $\partial/\partial a, \partial/\partial b, \partial/\partial c$ act on the operator only and not on the function $F(t,s,c)$; note that for an operator $A$, $(\partial A)F = [\partial, A]F$. Comparing the two ways (9.5) and (9.6) of computing (9.4), we obtain $[\mathbb{V}_k, \mathbb{U}_E] = 0$, with $\mathbb{V}_k$ as defined in the beginning of this section. The rest follows from the next argument:

$$[\mathbb{V}_\ell, \mathbb{U}_E^n] = \sum_{k=1}^n \mathbb{U}_E^{n-k}[\mathbb{V}_\ell, \mathbb{U}_E]\mathbb{U}_E^{k-1} = 0.$$

Incidentally, in view of Corollary 8.2, it also implies that $[\mathbb{V}_\ell, e^{-\lambda \mathbb{U}_E}] = 0$, since $e^{-\lambda \mathbb{U}_E} = \sum_{n=0}^\infty \frac{(-\lambda)^n}{n!}\mathbb{U}_E^n$. □

## 10. Vertex representation of probabilities and Virasoro constraints

Consider a weight $\rho(y,z)dy\,dz := \rho_{t,s}(y,z) := e^{V_{t,s}(y,z)}dy\,dz$ on $\mathbb{R}^2$, with $\rho_0 = e^{V_0}$, where

$$\tag{10.1} V_{t,s}(y,z) := cyz + \sum_1^\infty t_i y^i - \sum_1^\infty s_i z^i.$$

Given the space of Hermitean matrices $\mathcal{H}_N$, and given

spectrum $M_1 = \{x_1, \ldots, x_N\}$ and

spectrum $M_2 = \{y_1, \ldots, y_N\}$, with $M_1, M_2 \in \mathcal{H}_N$,

we define, for a set $E \subset \mathbb{R}^2$,

$$\mathcal{H}^2_{N,E} = \{(M_1, M_2) \in \mathcal{H}^2_N \text{ with all } (x_k, y_\ell) \in E\}.$$

Consider the product Haar measure $dM_1 dM_2$ on the product space $\mathcal{H}^2_N$, with each $dM_i$, decomposed into its radial part and its angular part, as in (0.16). Also define the probability measure

$$\frac{dM_1 dM_2 e^{\operatorname{Tr} V_{t,s}(M_1, M_2)}}{\int\int_{\mathcal{H}^2_N} dM_1 dM_2 e^{\operatorname{Tr} V_{t,s}(M_1, M_2)}}.$$

Recall from (9.3), the definition of the vector $\mathbb{J}^{(2)}_k$; also define another one $\tilde{\mathbb{J}}^{(2)}_k$:[14]

$$\tag{10.2} \mathbb{J}^{(2)}_k = (J^{(2)}_{k,n})_{n\in\mathbb{Z}} = \frac{1}{2}(J^{(2)}_k + (2n+k+1)J^{(1)}_k + n(n+1)J^{(0)}_k)_{n\in\mathbb{Z}},$$

$$\tilde{\mathbb{J}}^{(2)}_k = (\tilde{J}^{(2)}_{k,n})_{n\in\mathbb{Z}} = \frac{1}{2}(\tilde{J}^{(2)}_k + (2n+k+1)\tilde{J}^{(1)}_k + n(n+1)J^{(0)}_k)_{n\in\mathbb{Z}}.$$

Given the disjoint union

$$\tag{10.3} E = E_1 \times E_2 := \cup_{i=1}^r [a_{2i-1}, a_{2i}] \times \cup_{i=1}^s [b_{2i-1}, b_{2i}] \subset \mathbb{R}^2,$$

define the following integral:

$$\tag{10.4} \mathbb{U}_E := \int\int_E \mathbb{X}_{12}(x,y) \rho_0(x,y) dx dy,$$

of the vertex operator $\mathbb{X}_{12}$, defined in (5.2). The main theorem of this section is:

---

[14] $\tilde{\mathbb{J}}^{(i)}_k = \mathbb{J}^{(i)}_k \Big|_{t \mapsto -s}$, $i = 1, 2$.

THEOREM 10.1. *Given the set E, as in (10.3), the probability*

$P(\text{all } M_1\text{-eigenvalues} \in E_1 \text{ and all } M_2\text{-eigenvalues} \in E_2)$

$$\tag{10.5} = \frac{\int\int_{\mathcal{H}^2_{n,E}} dM_1 dM_2\, e^{\operatorname{Tr} V_{t,s}(M_1,M_2)}}{\int\int_{\mathcal{H}^2_n} dM_1 dM_2\, e^{\operatorname{Tr} V_{t,s}(M_1,M_2)}} =: \frac{\tau_n^E}{\tau_n}$$

*is a ratio of two $\tau$-functions $\tau_n^E$ and $\tau_n$, such that*

$$\tau_n^E = ((\mathbb{U}_E)^n \tau)_n.$$

*Moreover, $\tau_n$ and $\tau_n^E$ satisfy the partial differential equations, labeled for $k \geq -1$,*

$$\tag{10.6} \begin{aligned} \left(-\sum_{i=1}^r a_i^{k+1}\frac{\partial}{\partial a_i} + J_{k,n}^{(2)}\right)\tau_n^E + c\, p_{k+n}(\tilde{\partial}_t) p_n(-\tilde{\partial}_s)\tau_1^E \circ \tau_{n-1}^E &= 0 \\ \left(-\sum_{i=1}^s b_i^{k+1}\frac{\partial}{\partial b_i} + \tilde{J}_{k,n}^{(2)}\right)\tau_n^E + c\, p_n(\tilde{\partial}_t) p_{k+n}(-\tilde{\partial}_s)\tau_1^E \circ \tau_{n-1}^E &= 0. \end{aligned}$$

*Remark.* Whenever some $a_i$ or $b_i = \infty$, we must interpret: $a_i^{k+1}\frac{\partial}{\partial a_i}$ or $b_i^{k+1}\frac{\partial}{\partial b_i} \equiv 0$; in particular $\tau_n$ satisfies the same equations, but without the boundary terms.

The following proposition is due to [12], [8], [9]:

PROPOSITION 10.2.

$$\tag{10.7} \int_{\mathcal{U}(n)} dU\, e^{c\operatorname{Tr} xUyU^\top} = \frac{(2\pi)^{\frac{n(n-1)}{2}}}{n!}\frac{\det(e^{cx_iy_j})_{1\leq i,j\leq n}}{\Delta(x)\Delta(y)}.$$

PROPOSITION 10.3. *For $E = E_1 \times E_2 \subset \mathbb{R}^2$, the following holds:*

$$\tag{10.8} \iint_{\mathcal{H}^2_{N,E}} e^{c\operatorname{Tr}(M_1 M_2)} e^{\operatorname{Tr}\sum_1^\infty(t_i M_1^i - s_i M_2^i)} dM_1 dM_2$$

$$= \iint_{E^N} \prod_{k=1}^N (dx_k dy_k e^{\sum_{i=1}^\infty(t_i x_k^i - s_i y_k^i) + c\, x_k y_k}) \Delta_N(x)\Delta_N(y).$$

*Proof.* Consider a symmetric function $f(x,y) := f(x_1,\ldots,x_N; y_1,\ldots,y_N)$ in $y_1,\ldots,y_N$ for given $x_1,\ldots,x_N$; then we have, using the skew-symmetry of the Vandermonde $\Delta_N(y)$,

$$\int\int_{\mathbb{R}^{2N}} \Delta_N(x)\Delta_N(y) f(x,y) \det(e^{x_i y_j})_{1\leq i,j\leq N}\, dxdy$$

$$= \int\int_{\mathbb{R}^{2N}} \Delta_N(x) \sum_{\sigma\in\Pi_N}(-1)^\sigma \Delta_N(y) f(x,y) e^{\sum_1^N x_i y_{\sigma(i)}}\, dxdy$$

$$
\begin{aligned}
&= \iint_{\mathbb{R}^{2N}} \Delta_N(x) \sum_{\sigma \in \Pi_N} (-1)^\sigma \Delta_N(y_{\sigma^{-1}(1)}, \ldots, y_{\sigma^{-1}(N)}) \\
&\qquad \cdot f(x; y_{\sigma^{-1}(1)}, \ldots, y_{\sigma^{-1}(N)}) e^{\sum_1^N x_i y_i} dx\, dy \\
&= \iint_{\mathbb{R}^{2N}} \Delta_N(x) \left( \sum_{\sigma \in \Pi_N} (-1)^\sigma (-1)^\sigma \right) \Delta_N(y_1, \ldots, y_N) f(x,y) e^{\sum_1^N x_i y_i} dx\, dy \\
&= N! \iint_{\mathbb{R}^{2N}} \Delta_N(x) \Delta_N(y) f(x,y) e^{\sum_1^N x_i y_i} dx\, dy.
\end{aligned}
$$

The function

$$
f(x,y) = \prod_1^N I_{E_1 \times E_2}(x_i, y_i) = \prod_1^N I_{E_1}(x_i) I_{E_2}(y_i)
$$

has the desired symmetry property, so that, using (10.7), one computes

$$
\iint_{\mathcal{H}_{N,E}^2} e^{c\operatorname{Tr}(M_1 M_2)} e^{\operatorname{Tr} \sum_1^\infty (t_i M_1^i - s_i M_2^i)} dM_1\, dM_2
$$

$$
\begin{aligned}
&= \iint_{E^N} \prod_1^N dx_i \prod_1^N dy_i \Delta_N^2(x) \Delta_N^2(y) \prod_{k=1}^N e^{\sum_{i=1}^\infty (t_i x_k^i - s_i y_k^i)} \\
&\qquad \cdot \iint_{\mathcal{U}(N)^2} dU_1\, dU_2\, e^{c\operatorname{Tr} U_1 x U_1^\top U_2 y U_2^\top} \\
&= \iint_{E^N} \prod_1^N dx_i \prod_1^N dy_i \Delta_N^2(x) \Delta_N^2(y) \prod_{k=1}^N e^{\sum_{i=1}^\infty (t_i x_k^i - s_i y_k^i)} \\
&\qquad \cdot \iint_{\mathcal{U}(N)^2} dU_1\, dU_2\, e^{c\operatorname{Tr} x U_2 y U_2^\top},
\end{aligned}
$$

$$\text{substituting } U_2 \text{ for } U_1^\top U_2,$$

$$
\begin{aligned}
&= c_N \iint_{\mathbb{R}^{2N}} \prod_1^N dx_i \prod_1^N dy_i \Delta_N^2(x) \Delta_N^2(y) f(x,y) \\
&\qquad \cdot \prod_{k=1}^N e^{\sum_{i=1}^\infty (t_i x_k^i - s_i y_k^i)} \frac{\det(e^{c x_i y_j})_{1 \le i,j \le N}}{\Delta_N(x) \Delta_N(y)} \\
&= c'_N \iint_{E^N} \prod_1^N dx_i \prod_1^N dy_i \Delta_N(x) \Delta_N(y) \prod_{k=1}^N e^{\sum_{i=1}^\infty (t_i x_k^i - s_i y_k^i) + c\, x_k y_k},
\end{aligned}
$$

where the last identity follows from the previous calculation. □

Consider now a more general weight $\rho(y,z)dydz := \rho_{t,s}(y,z)dydz := e^{V_{t,s}(y,z)}dydz$ on $\mathbb{R}^2$, with $\rho_0 = e^{V_0}$, where

$$(10.9) \quad V_{t,s}(y,z) := V_0(y,z) + \sum_1^\infty t_i y^i - \sum_1^\infty s_i z^i = \sum_{i,j\geq 1} c_{ij} y^i z^j + \sum_1^\infty t_i y^i - \sum_1^\infty s_i z^i,$$

with arbitrary $V_0$ and the inner product with regard to a subset $E \subset \mathbb{R}^2$

$$(10.10) \quad \langle f,g\rangle_E = \int_E dy\, dz \rho_{t,s}(y,z) f(y) g(z).$$

Given the moment matrix (over $E$),

$$(10.11) \quad m_n(t,s,c) =: (\mu_{ij})_{0\leq ij \leq n-1} = (\langle y^i, z^j\rangle_E)_{0\leq i,j\leq n-1},$$

according to [2], [3], the Borel decomposition of the semi-infinite matrix[15]

$$m_\infty = S_1^{-1} S_2 \text{ with } S_1 \in \mathcal{D}_{-\infty,0}, \quad S_2 \in \mathcal{D}_{0,\infty}$$

with $S_1$ having 1's on the diagonal, and $S_2$ having $h_i$'s on the diagonal, leads to two strings $(p^{(1)}(y), p^{(2)}(z))$ of monic polynomials in one variable (dependent on $E$), constructed, in terms of the character $\bar\chi(z) = (z^n)_{n\in\mathbb{Z}, n\geq 0}$, as follows:

$$(10.12) \quad p^{(1)}(y) =: S_1 \bar\chi(y), \quad p^{(2)}(z) =: h(S_2^{-1})^\top \bar\chi(z).$$

We call these two sequences *bi-orthogonal polynomials*; in fact, according to [3] the Borel decomposition of $m_\infty = S_1^{-1} S_2$ above is equivalent to the "orthogonality" relations of the polynomials:

$$(10.13) \quad \langle p_n^{(1)}, p_m^{(2)}\rangle_E = \delta_{n,m} h_n.$$

The matrices
$$L_1 := S_1 \Lambda S_1^{-1}, \text{ and } L_2 := S_2 \Lambda^\top S_2^{-1},$$

interact with the vector of string orthogonal polynomials, as follows:

$$(10.14) \quad L_1 p^{(1)}(y) = y p^{(1)}(y), \quad h L_2^\top h^{-1} p^{(2)}(z) = z p^{(2)}(z).$$

Also define vectors $\Psi_1$ and $\Psi_2^*$, as follows:

$$(10.15) \quad \begin{aligned} \Psi_1(z) &:= e^{\Sigma t_k z^k} p^{(1)}(z) \quad \text{and} & \Psi_2^*(z) &:= e^{-\Sigma s_k z^{-k}} h^{-1} p^{(2)}(z^{-1}) \\ &= e^{\Sigma t_k z^k} S_1 \bar\chi(z) & &= e^{-\Sigma s_k z^{-k}} (S_2^{-1})^\top \bar\chi(z^{-1}). \end{aligned}$$

---

[15] $\mathcal{D}_{k,\ell}$ ($k < \ell \in \mathbb{Z}$) denotes the set of band matrices with zeros outside the strip $(k,\ell)$.

As a function of $(t,s)$, the couple $L := (L_1, L_2)$ satisfies the two-Toda lattice equations (2.3), and $\Psi_1$ and $\Psi_2^*$ satisfy the equations (2.7); remember that $L$, $\Psi_1$ and $\Psi_2^*$ all depend on $E$.

Moreover, according to [2], Theorem 3.4, the determinant of the moment matrix can be expressed as a $2n$-uple integral over $E^n \subset \mathbb{R}^{2n}$:

(10.16)
$$\begin{aligned} n! \det m_n(t,s,c) &= \iint_{(u,v) \in E^n \subseteq \mathbb{R}^{2n}} \Delta_n(u) \Delta_n(v) \prod_{k=1}^n \left( e^{V_{t,s}(u_k, v_k)} du_k dv_k \right) \\ &= n! \det \left( E_n(t) m_\infty(0,0,c) E_n(-s)^\top \right) \\ &= \prod_0^{n-1} h_i(t,s,c) \\ &= \tau_n^E(t,s,c), \end{aligned}$$

where $E_n(t) :=$ (the first $n$ rows of $e^{\sum_1^\infty t_n \Lambda^n}$) is a matrix of Schur polynomials $p_n(t)$. Also $\tau_n(t,s,c)$ is a $\tau$-function with regard to $t$ and $s$. Note that for $V_0 = cxy$ and $E = E_1 \times E_2$, this integral is precisely the one obtained in Proposition 10.3.

PROPOSITION 10.4. *Given the bi-orthogonal polynomials $(p_k^{(1)}, p_k^{(2)})$ for a general weight $\rho_0 = e^{V_0}$ on $\mathbb{R}^2$, the kernel defined in terms of (10.15),*

(10.17)
$$\begin{aligned} K(y,y';z,z') := K_n(y,z') &:= \sum_{0 \leq k < n} \Psi_{1,k}(y) \Psi_{2,k}^*(z'^{-1}) \\ &= \sum_{0 \leq k < n} e^{\sum_1^\infty t_i y^i} p_k^{(1)}(y) h_k^{-1} p_k^{(2)}(z') e^{-\sum_1^\infty s_i z'^i}, \end{aligned}$$

*defines a projector; i.e., it has the reproducing property with regard to the measure $\rho_0 dz \, dz'$:*

$$\iint_{\mathbb{R}^2} K(y,y';z,z') K(z,z';u,u') \rho_0(z,z') dz \, dz' = K(y,y';u,u')$$

*and*

(10.18)
$$\iint_{\mathbb{R}^2} K(z,z';z,z') \rho_0(z,z') dz \, dz' = n.$$

*Proof.* Using the explicit expression (10.17), using the fact that $\rho_{t,s}(y,z) = \rho_0(y,z) e^{\sum_1^\infty (t_i y^i - s_i z^i)}$ in the second equality, and using the orthogonality relation (10.13), in the third equality, one computes

$$\iint_{\mathbb{R}^2} K(y, y'; z, z') K(z, z'; u, u') \rho_0(z, z') dz dz'$$

$$= \iint_{\mathbb{R}^2} \left( \sum_{0 \leq k < n} e^{\Sigma t_i y^i} p_k^{(1)}(y) h_k^{-1} p_k^{(2)}(z') e^{-\Sigma s_i z'^i} \right)$$

$$\cdot \left( \sum_{0 \leq \ell < n} e^{\Sigma t_i z^i} p_\ell^{(1)}(z) h_\ell^{-1} p_\ell^{(2)}(u') e^{-\Sigma s_i u'^i} \right) \rho_0(z, z') dz\, dz'$$

$$= \iint_{\mathbb{R}^2} \sum_{0 \leq k, \ell < n} e^{\Sigma t_i y^i} p_k^{(1)}(y) h_k^{-1} p_\ell^{(2)}(u')$$

$$\cdot e^{-\Sigma s_i u'^i} p_k^{(2)}(z') p_\ell^{(1)}(z) h_\ell^{-1} \rho_{t,s}(z, z') dz\, dz'$$

$$= \sum_{0 \leq k < n} e^{\Sigma t_i y^i} p_k^{(1)}(y) h_k^{-1} p_k^{(2)}(u') e^{-\Sigma s_i u'^i} = K_n(y, u') = K(y, y'; u, u'),$$

and

$$\iint_{\mathbb{R}^2} K(z, z'; z, z') \rho_0(z, z') dz\, dz'$$

$$= \iint_{\mathbb{R}^2} K_n(z, z') \rho_0(z, z') dz\, dz'$$

$$= \sum_{0 \leq k < n} \iint_{\mathbb{R}^2} p_k^{(1)}(z) p_k^{(2)}(z') h_k^{-1} \rho_{t,s}(z, z') dz\, dz'$$

$$= n. \qquad \square$$

Consider, for $E \in \mathbb{R}^2$, the vectors (see (10.16))

$$(10.19) \quad \tau^E = \left( \tau_n^E \right) := \left( \iint_{E^n} \prod_{k=1}^n (dx_k dy_k \rho_{t,s}(x_k, y_k)) \Delta_n(x) \Delta_n(y) \right)_{n \geq 0},$$

$$\tau := (\tau_n)_{n \geq 0} := \left( \tau_n^{\mathbb{R}^2} \right)_{n \geq 0},$$

with $\rho_0$ as in (10.9), and the vector of operators $\mathbb{U}_E$, defined in (10.4), but for the weight $\rho_0$ as in (10.9). Given the set $E \subset \mathbb{R}^2$, define, in accordance with (9.3):

$$(10.20) \quad \mathbb{V}_k := -\sum_{i=1}^r a_i^{k+1} \frac{\partial}{\partial a_i} + \mathbb{J}_k^{(2)} + \sum_{i,j \geq 1} i c_{ij} \frac{\partial}{\partial c_{i+k,j}},$$

$$(10.21) \quad \tilde{\mathbb{V}}_k := -\sum_{i=1}^s b_i^{k+1} \frac{\partial}{\partial b_i} + \tilde{\mathbb{J}}_k^{(2)} + \sum_{i,j \geq 1} j c_{ij} \frac{\partial}{\partial c_{i,j+k}}.$$

We now state:

PROPOSITION 10.5. *For $E = E_1 \times E_2 \subset \mathbb{R}^2$,[16] we have*

$$\tau_n^E = ((\mathbb{U}_E)^n \tau)_n. \tag{10.22}$$

*Proof.* In what follows, we use the monic bi-orthogonal polynomials $p_i^{(1)}, p_j^{(2)}$, defined by $\rho_{t,s}(x, y)$ on $\mathbb{R}^2$; therefore the $h_i(t, s, c)$ are the $\mathbb{R}^2$ inner products. We first compute, using (10.16) for $E = \mathbb{R}^2$ and Proposition 10.3, and remembering notation (10.4), and formulae (10.16) and (10.17):

$$\begin{aligned}
\frac{\tau_n^E}{\tau_n} &= \left(\prod_0^{n-1} h_i^{-1}\right) \iint_{E^n} \prod_{k=1}^n (dx_k dy_k \rho_{t,s}(x_k, y_k)) \Delta_n(\vec{x}) \Delta_n(\vec{y}) \\
&= \left(\prod_0^{n-1} h_i^{-1}\right) \iint_{E^n} \prod_{k=1}^n (dx_k dy_k \rho_{t,s}(x_k, y_k)) \det(p_{i-1}^{(1)}(x_j))_{1 \leq i,j \leq n} \det(p_{i-1}^{(2)}(y_j))_{1 \leq i,j \leq n} \\
&= \iint_{E^n} \prod_{k=1}^n (dx_k dy_k \rho_0(x_k, y_k)) \det \left( e^{\Sigma t_i x_k^i} \sum_{i=1}^n p_{i-1}^{(1)}(x_k) h_{i-1}^{-1} p_{i-1}^{(2)}(y_\ell) e^{-\Sigma s_i y_\ell^i} \right)_{1 \leq k,\ell \leq n} \\
&= \iint_{E^n} \prod_{k=1}^n (dx_k dy_k \rho_0(x_k, y_k)) \det \left( \sum_{0 \leq i \leq n-1} \Psi_{1i}(x_k) \Psi_{2i}^*(y_\ell^{-1}) \right)_{1 \leq k,\ell \leq n} \\
&= \iint_{E^n} \prod_{k=1}^n (\rho_0 dx_k dy_k) \det(K_n(x_k, y_\ell))_{1 \leq k,\ell \leq n} \\
&= \iint_{E^n} \prod_{k=1}^n (\rho_0 dx_k dy_k) \left(\frac{1}{\tau} \prod_{k=1}^n \mathbb{X}_{12}(x_k, y_k) \tau \right)_n, \quad \text{using (8.1)}, \\
&= \left(\frac{1}{\tau} \left(\iint_E \mathbb{X}_{12}(x, y) \rho_0(x, y) dx dy\right)^n \tau \right)_n,
\end{aligned}$$

establishing (10.22). □

*Proof of Theorem* 10.1. In [2] (see for instance the introduction), we have shown that the vector $\tau = \tau^{\mathbb{R}^2}$, which is independent of the $a_i$'s and $b_i$'s, satisfies the infinite set of equations, for $k \geq -1$,

$$\begin{aligned}
\mathbb{V}_k \tau &= (\mathbb{J}_k^{(2)} + \sum_{i,j \geq 1} i c_{ij} \frac{\partial}{\partial c_{k+i,j}}) \tau = 0 \\
\tilde{\mathbb{V}}_k \tau &= (\tilde{\mathbb{J}}_k^{(2)} + \sum_{i,j \geq 1} j c_{ij} \frac{\partial}{\partial c_{i,j+k}}) \tau = 0.
\end{aligned}$$

---

[16]$((\mathbb{U}_E)^n \tau)_n$ means the $n^{\text{th}}$ component of the vector $(\mathbb{U}_E)^n \tau$.

According to Theorem 9.1 , we also have $[\mathbb{V}_k, (\mathbb{U}_E)^n] = 0$ and thus

$$0 = [\mathbb{V}_k, (\mathbb{U}_E)^n]\tau = \mathbb{V}_k(\mathbb{U}_E)^n\tau - (\mathbb{U}_E)^n\mathbb{V}_k\tau = \mathbb{V}_k(\mathbb{U}_E)^n\tau;$$

taking the $n^{\text{th}}$ component, we find $(\mathbb{V}_k(\mathbb{U}_E)^n\tau)_n = 0$ and similarly with $\mathbb{V}_k$ replaced by $\tilde{\mathbb{V}}_k$. Since $((\mathbb{U}_E)^n\tau)_n = \tau_n^E$ by (10.22), this leads to:

$$(10.23) \quad \left(-\sum_{i=1}^r a_i^{k+1}\frac{\partial}{\partial a_i} + J_{k,n}^{(2)} + \sum_{i,j\geq 1} ic_{ij}\frac{\partial}{\partial c_{k+i,j}}\right)\tau_n^E = 0$$

$$\left(-\sum_{i=1}^s b_i^{k+1}\frac{\partial}{\partial b_i} + \tilde{J}_{k,n}^{(2)} + \sum_{i,j\geq 1} jc_{ij}\frac{\partial}{\partial c_{i,j+k}}\right)\tau_n^E = 0.$$

But, by p. 285 of [2] and by Theorem 6.1,

$$(10.24) \quad \frac{\partial \tau_n^E}{\partial c_{\alpha\beta}} = \tau_n^E \sum_{i=0}^{n-1}(L_1^\alpha L_2^\beta)_{ii} = p_{\alpha+n-1}(\tilde{\partial}_t)p_{\beta+n-1}(-\tilde{\partial}_s)\tau_1^E \circ \tau_{n-1}^E.$$

Remember, one is really interested in the probability, expressed by $\tau$-functions (see (10.8) and (10.16)),

$$P((M_1, M_2) \in \mathcal{H}_{n,E}^2) = \frac{\iint_{\mathcal{H}_{n,E}^2} dM_1 dM_2\, e^{\operatorname{Tr} V_{t,s}(M)}}{\iint_{\mathcal{H}_n^2} dM_1 dM_2\, e^{\operatorname{Tr} V_{t,s}(M)}} = \frac{\tau_n^E}{\tau_n},$$

for $\rho_0 = e^{V_0} = e^{cxy}$; thus, we must set all $c_{ij} = 0$, but $c = c_{11}$; this leads to the statements (10.6), ending the proof of Theorem 10.1. □

## 11. PDE's for the joint statistics of the spectra of Gaussian coupled random matrices

Consider the Gaussian probability measure

$$(11.1) \quad c_n dM_1 dM_2 e^{-\frac{1}{2}\operatorname{Tr}(M_1^2 + M_2^2 - 2cM_1M_2)},$$

defined over the space of Hermitean matricex $\mathcal{H}_n^2 = \mathcal{H}_n \times \mathcal{H}_n$, with a coupling constant $c$. Consider the joint probability

$$(11.2) \quad P_n(E) := P(\text{all } (M_1\text{-eigenvalues}) \in E_1, (M_2\text{-eigenvalues}) \in E_2)$$

for a set of the form $E = E_1 \times E_2 := \cup_{i=1}^r [a_{2i-1}, a_{2i}] \times \cup_{i=1}^s [b_{2i-1}, b_{2i}] \subset \mathbb{R}^2$. Before stating the theorem, we remind the reader of the differential operators $\mathcal{A}_k, \mathcal{B}_k$, depending on the boundary points of $E$ and the coupling constant $c$:

$$(11.3)$$
$$\mathcal{A}_1 = \frac{1}{c^2-1}\left(\sum_1^r \frac{\partial}{\partial a_j} + c\sum_1^s \frac{\partial}{\partial b_j}\right), \quad \mathcal{B}_1 = -\frac{1}{c^2-1}\left(c\sum_1^r \frac{\partial}{\partial a_j} + \sum_1^s \frac{\partial}{\partial b_j}\right),$$
$$\mathcal{A}_2 = \sum_{j=1}^r a_j\frac{\partial}{\partial a_j} - c\frac{\partial}{\partial c}, \quad \mathcal{B}_2 = \sum_{j=1}^s b_j\frac{\partial}{\partial b_j} - c\frac{\partial}{\partial c};$$

they form a Lie algebra parametrized by $c$:

(11.4) $\quad [\mathcal{A}_1, \mathcal{B}_1] = 0 \quad [\mathcal{A}_1, \mathcal{A}_2] = \dfrac{1+c^2}{1-c^2}\mathcal{A}_1 \quad [\mathcal{A}_2, \mathcal{B}_1] = \dfrac{2c}{1-c^2}\mathcal{A}_1$

$\qquad\qquad [\mathcal{A}_2, \mathcal{B}_2] = 0 \quad [\mathcal{A}_1, \mathcal{B}_2] = \dfrac{-2c}{1-c^2}\mathcal{B}_1 \quad [\mathcal{B}_1, \mathcal{B}_2] = \dfrac{1+c^2}{1-c^2}\mathcal{B}_1.$

We now prove Theorem 0.5, as announced in the introduction:

THEOREM 11.1 (Gaussian probability). *The joint statistics* (11.2) *satisfies the nonlinear third-order partial differential equation*[17] ($F_n := \frac{1}{n}\log P_n(E)$):

(11.5)
$$\left\{\mathcal{B}_2\mathcal{A}_1 F_n \,,\, \mathcal{B}_1\mathcal{A}_1 F_n + \frac{c}{c^2-1}\right\}_{\mathcal{A}_1} - \left\{\mathcal{A}_2\mathcal{B}_1 F_n \,,\, \mathcal{A}_1\mathcal{B}_1 F_n + \frac{c}{c^2-1}\right\}_{\mathcal{B}_1} = 0.$$

*Remark.* When $E_1 = E_2$, equation (11.5) is trivially satisfied.

*Proof.* From (10.5), it clearly follows that

$$P_n(E) = \left.\frac{\tau_n^E(t,s,c_{ij})}{\tau_n^{\mathbb{R}^2}(t,s,c_{ij})}\right|_{\mathcal{L}},$$

where $\tau_n^E$ is an integral over $E^n \subset \mathbb{R}^{2n}$, i.e., $(x,y) \in E_1^n \times E_2^n = E^n$,

(11.6)
$$\tau_n^E(t,s,c_{ij}) = \iint_{E^n} dxdy \Delta_n(x)\Delta_n(y)$$
$$\cdot \prod_{k=1}^n e^{-\frac{1}{2}(x_k^2+y_k^2-2cx_ky_k)+\sum_{i=1}^\infty(t_ix_k^i-s_iy_k^i)+\sum_{\substack{i,j\geq 1\\(i,j)\neq(1,1)}} c_{ij}x_k^i y_k^j},$$

and where $\mathcal{L}$ denotes the locus

$$\mathcal{L} = \{t_i = s_i = 0,\ c_{11} = c \text{ and all other } c_{ij} = 0\}.$$

Observe the following involution on $\tau_n^E(t,s) = \tau_n(t,s,a,b,c)$:

(11.7) $\qquad\qquad \tau_n(-s,-t,b,a,c) = \tau_n(t,s,a,b,c),$

implying for the $\mathcal{A}_i$, $\mathcal{B}_i$, defined in (11.3) and $\mathcal{V}_i$, $\mathcal{W}_i$, defined below:

$$\mathcal{A}_i \leftrightarrow (-1)^i \mathcal{B}_i, \quad \mathcal{V}_i \leftrightarrow (-1)^i \mathcal{W}_i.$$

In view of (11.6), we write down the Virasoro equations (10.23) for $\tau_n^E$, but with the shifts $t_2 \mapsto -\frac{1}{2} + t_2$, $s_2 \mapsto \frac{1}{2} + s_2$. It is convenient to consider new Virasoro generators $\mathcal{V}_k$ and $\mathcal{W}_k$, such that

(11.8) $\qquad\qquad \mathcal{V}_k|_{\mathcal{L}} = \pm\dfrac{\partial}{\partial t_k} \quad \text{and} \quad \mathcal{W}_k|_{\mathcal{L}} = \pm\dfrac{\partial}{\partial s_k},$

---

[17] in terms of the Wronskian $\{f,g\}_X = Xf.g - f.Xg$, with regard to a first-order differential operator $X$.

namely, in terms of (10.20) and (10.21):

(11.9)
$$\begin{aligned}
\mathcal{V}_1 &= \frac{1}{c^2-1}(\mathbb{V}_{-1} + c\tilde{\mathbb{V}}_{-1}) \\
&= \frac{\partial}{\partial t_1} + \frac{n(t_1 - cs_1)}{c^2-1} \\
&\quad + \frac{1}{c^2-1}\left(\sum_{i\geq 2} i(t_i\frac{\partial}{\partial t_{i-1}} + cs_i\frac{\partial}{\partial s_{i-1}}) + \sum_{\substack{i,j\geq 1 \\ i,j\neq(1,1)}} c_{ij}(i\frac{\partial}{\partial c_{i-1,j}} + jc\frac{\partial}{\partial c_{i,j-1}})\right) \\
\mathcal{W}_1 &= \frac{1}{1-c^2}(c\mathbb{V}_{-1} + \tilde{\mathbb{V}}_{-1}) \\
&= \frac{\partial}{\partial s_1} - \frac{n(ct_1 - s_1)}{c^2-1} \\
&\quad - \frac{1}{c^2-1}\left(\sum_{i\geq 2} i(ct_i\frac{\partial}{\partial t_{i-1}} + s_i\frac{\partial}{\partial s_{i-1}}) + \sum_{\substack{i,j\geq 1 \\ i,j\neq(1,1)}} c_{ij}(ci\frac{\partial}{\partial c_{i-1,j}} + j\frac{\partial}{\partial c_{i,j-1}})\right), \\
\mathcal{V}_2 &:= \mathbb{V}_0 - c\frac{\partial}{\partial c} \\
&= -\frac{\partial}{\partial t_2} + \sum_{i\geq 1} it_i\frac{\partial}{\partial t_i} + \frac{n(n+1)}{2} + \sum_{\substack{i,j\geq 1 \\ (i,j)\neq(1,1)}} ic_{ij}\frac{\partial}{\partial c_{ij}} \\
\mathcal{W}_2 &:= \tilde{\mathbb{V}}_0 - c\frac{\partial}{\partial c} \\
&= \frac{\partial}{\partial s_2} + \sum_{i\geq 1} is_i\frac{\partial}{\partial s_i} + \frac{n(n+1)}{2} + \sum_{\substack{i,j\geq 1 \\ (i,j)\neq(1,1)}} jc_{ij}\frac{\partial}{\partial c_{ij}}.
\end{aligned}$$

With this new notation, and by virtue of (10.23) and (10.2), the $\tau_n$'s satisfy for all $n \geq 1$:

(11.10) $\qquad \mathcal{A}_k\tau_n = \mathcal{V}_k\tau_n \quad \text{and} \quad \mathcal{B}_k\tau_n = \mathcal{W}_k\tau_n, \quad k = 1, 2$.

In particular, on the locus $\mathcal{L}$, we have from (11.9),

$$\mathcal{A}_1\tau_n\Big|_{\mathcal{L}} = \frac{\partial \tau_n}{\partial t_1}\Big|_{\mathcal{L}} \qquad\qquad \mathcal{B}_1\tau_n\Big|_{\mathcal{L}} = \frac{\partial \tau_n}{\partial s_1}\Big|_{\mathcal{L}}$$

$$\mathcal{A}_2\tau_n\Big|_{\mathcal{L}} = \left(-\frac{\partial}{\partial t_2} + \frac{n(n+1)}{2}\right)\tau_n\Big|_{\mathcal{L}} \quad \mathcal{B}_2\tau_n\Big|_{\mathcal{L}} = \left(\frac{\partial}{\partial s_2} + \frac{n(n+1)}{2}\right)\tau_n\Big|_{\mathcal{L}},$$

and so
(11.11)
$$\frac{\partial}{\partial t_1}\log\tau_n\Big|_{\mathcal{L}} = \mathcal{A}_1\log\tau_n\Big|_{\mathcal{L}} \qquad\qquad \frac{\partial}{\partial s_1}\log\tau_n\Big|_{\mathcal{L}} = \mathcal{B}_1\log\tau_n\Big|_{\mathcal{L}}$$

$$\frac{\partial}{\partial t_2}\log\tau_n\Big|_{\mathcal{L}} = -\mathcal{A}_2\log\tau_n\Big|_{\mathcal{L}} + \frac{n(n+1)}{2} \quad \frac{\partial}{\partial s_2}\log\tau_n\Big|_{\mathcal{L}} = \mathcal{B}_2\log\tau_n\Big|_{\mathcal{L}} - \frac{n(n+1)}{2}.$$

Using $[\mathcal{B}_1, \mathcal{V}_1]\big|_{\mathcal{L}} = 0$, we have

$$\begin{aligned}
\mathcal{B}_1 \mathcal{A}_1 \tau_n \big|_{\mathcal{L}} &= \mathcal{B}_1 \mathcal{V}_1 \tau_n \big|_{\mathcal{L}} \\
&= \mathcal{V}_1 \mathcal{B}_1 \tau_n \big|_{\mathcal{L}} \\
&= \mathcal{V}_1 \mathcal{W}_1 \tau_n \big|_{\mathcal{L}} \\
&= \frac{\partial}{\partial t_1} \left( \frac{\partial}{\partial s_1} + n \frac{ct_1 - s_1}{1 - c^2} \right) \tau_n \bigg|_{\mathcal{L}} \\
&= \left( \frac{\partial^2}{\partial t_1 \partial s_1} + \frac{nc}{1 - c^2} \right) \tau_n \bigg|_{\mathcal{L}},
\end{aligned}$$

and so, on the locus $\mathcal{L}$,

(11.12) $$\frac{\partial^2}{\partial t_1 \partial s_1} \log \tau_n \bigg|_{\mathcal{L}} = \mathcal{B}_1 \mathcal{A}_1 \log \tau_n + \frac{nc}{c^2 - 1}.$$

Using $[\mathcal{B}_2, \mathcal{V}_1]\big|_{\mathcal{L}} = 0$, we see that

$$\begin{aligned}
\mathcal{B}_2 \mathcal{A}_1 \tau_n \big|_{\mathcal{L}} &= \mathcal{B}_2 \mathcal{V}_1 \tau_n \big|_{\mathcal{L}} \\
&= \mathcal{V}_1 \mathcal{B}_2 \tau_n \big|_{\mathcal{L}} \\
&= \mathcal{V}_1 \mathcal{W}_2 \tau_n \big|_{\mathcal{L}} \\
&= \frac{\partial}{\partial t_1} \left( \frac{\partial}{\partial s_2} + \frac{n(n+1)}{2} \right) \tau_n \bigg|_{\mathcal{L}} \\
&= \left( \frac{\partial^2}{\partial t_1 \partial s_2} + \frac{n(n+1)}{2} \frac{\partial}{\partial t_1} \right) \tau_n \bigg|_{\mathcal{L}},
\end{aligned}$$

and so, on $\mathcal{L}$, we have[18]

(11.13) $$\frac{\partial^2}{\partial t_1 \partial s_2} \log \tau_n \bigg|_{\mathcal{L}} = \mathcal{B}_2 \mathcal{A}_1 \log \tau_n.$$

Setting (11.11), (11.12), (11.13) into the formula of Proposition 3.3 (for $k = 2$, as spelled out in Lemma 3.4) and its dual, namely

(11.14) $$-\frac{\partial}{\partial s_1} \log \frac{\tau_{n+1}}{\tau_{n-1}} = \frac{\frac{\partial^2}{\partial t_1 \partial s_2} \log \tau_n}{\frac{\partial^2}{\partial t_1 \partial s_1} \log \tau_n} \quad \text{and} \quad \frac{\partial}{\partial t_1} \log \frac{\tau_{n+1}}{\tau_{n-1}} = \frac{\frac{\partial^2}{\partial s_1 \partial t_2} \log \tau_n}{\frac{\partial^2}{\partial s_1 \partial t_1} \log \tau_n},$$

---

[18]Using the following relation for non-commutative operators $X$ and $Y$
$$XY \log f = \frac{1}{f^2} \left( fXYf - Xf\, Yf \right).$$

one is led to an expression for $\mathcal{B}_1 \log \frac{\tau_{n+1}}{\tau_{n-1}}$ and, using the involution (11.7), a dual expression for $\mathcal{A}_1 \log \frac{\tau_{n+1}}{\tau_{n-1}}$:

$$
\begin{aligned}
-\mathcal{A}_1 \log \frac{\tau_{n+1}}{\tau_{n-1}} &= \frac{\mathcal{A}_2 \mathcal{B}_1 \log \tau_n}{\mathcal{A}_1 \mathcal{B}_1 \log \tau_n + \frac{nc}{c^2-1}} \\
-\mathcal{B}_1 \log \frac{\tau_{n+1}}{\tau_{n-1}} &= \frac{\mathcal{B}_2 \mathcal{A}_1 \log \tau_n}{\mathcal{B}_1 \mathcal{A}_1 \log \tau_n + \frac{nc}{c^2-1}}.
\end{aligned}
\tag{11.15}
$$

Upon taking $\mathcal{A}_1$ of the second expression, subtracting from it $\mathcal{B}_1$ of the first one and using $[\mathcal{A}_1, \mathcal{B}_1] = 0$, one finds the following identity

$$
\mathcal{A}_1 \frac{\mathcal{B}_2 \mathcal{A}_1 \log \tau_n}{\mathcal{B}_1 \mathcal{A}_1 \log \tau_n + \frac{nc}{c^2-1}} - \mathcal{B}_1 \frac{\mathcal{A}_2 \mathcal{B}_1 \log \tau_n}{\mathcal{A}_1 \mathcal{B}_1 \log \tau_n + \frac{nc}{c^2-1}} = 0.
$$

This difference amounts to the equality of two Wronskians ($G_n := \frac{1}{n} \log \tau_n$):

$$
\left\{ \mathcal{B}_2 \mathcal{A}_1 G_n \,,\, \mathcal{B}_1 \mathcal{A}_1 G_n + \frac{c}{c^2-1} \right\}_{\mathcal{A}_1} = \left\{ \mathcal{A}_2 \mathcal{B}_1 G_n \,,\, \mathcal{A}_1 \mathcal{B}_1 G_n + \frac{c}{c^2-1} \right\}_{\mathcal{B}_1}.
\tag{11.16}
$$

Because of the fact that

$$
\log P_n(E) = \log(\tau_n(E)/\tau_n(\mathbb{R}^2)) = \log \tau_n(E) - \log \tau_n(\mathbb{R}^2),
$$

together with the fact that $\mathcal{A}_1 \tau_n(\mathbb{R}^2) = \mathcal{B}_1 \tau_n(\mathbb{R}^2) = 0$, we have that $F_n(E) := \frac{1}{n} \log P_n(E)$ satisfies (11.16) as well, thus leading to (11.5). □

*Remark.* For small $n$, the equation (11.5) for $\tau_n = \det m_n$ can be checked, using the explicit moment matrix $m_n = (\mu_{ij})_{0 \le i,j \le n-1}$, where

$$
\mu_{ij} = \int_{E_1} dx \int_{E_2} dy\, x^i y^j e^{-\frac{1}{2}(x^2+y^2-2cxy)}.
$$

It suffices to compute the action of $\mathcal{A}_i$ and $\mathcal{B}_i$ on $\mu_{ij}$, namely

$$
\mathcal{A}_1 \mu_{ij} = \mu_{i+1,j} + \frac{i\mu_{i-1,j} + cj\mu_{i,j-1}}{c^2-1} \qquad \mathcal{A}_2 \mu_{ij} = (i+1)\mu_{ij} - \mu_{i+2,j}
\tag{11.17}
$$

$$
\mathcal{B}_1 \mu_{ij} = -\mu_{i,j+1} + \frac{j\mu_{i,j-1} + ci\mu_{i-1,j}}{1-c^2} \qquad \mathcal{B}_2 \mu_{ij} = (j+1)\mu_{ij} - \mu_{i,j+2},
$$

and check equation (11.15), at least for small $n$.

## 12. Coupled random matrices with the Laguerre statistics

Consider the Laguerre probability measure

$$
c_n dM_1 dM_2 e^{\operatorname{Tr}(-M_1 + \alpha \log M_1 - M_2 + \alpha \log M_2 + cM_1 M_2)},
\tag{12.1}
$$

defined over the space of Hermitean matricex $\mathcal{H}_n^2 = \mathcal{H}_n \times \mathcal{H}_n$, with a coupling constant $c$. Consider the joint probability

(12.2)    $P_n(E) := P(\text{all } (M_1\text{-eigenvalues}) \in E_1, (M_2\text{-eigenvalues}) \in E_2)$

for a set of the form $E = E_1 \times E_2 := \cup_{i=1}^{r}[a_{2i-1}, a_{2i}] \times \cup_{i=1}^{s}[b_{2i-1}, b_{2i}] \subset \mathbb{R}^2$. Before stating the theorem, we remind the reader of the differential operators $\mathcal{A}_k, \mathcal{B}_k$, depending on the boundary points of $E$ and the coupling constant $c$:

(12.3)
$$\mathcal{A}_1 = \sum_{j=1}^{r} a_j \frac{\partial}{\partial a_j} - c\frac{\partial}{\partial c}, \qquad \mathcal{B}_1 = \sum_{j=1}^{s} b_j \frac{\partial}{\partial b_j} - c\frac{\partial}{\partial c},$$
$$\mathcal{A}_2 = \sum_{j=1}^{r} a_j^2 \frac{\partial}{\partial a_j} + (n+1+\alpha)\mathcal{A}_1 - c\frac{\partial}{\partial c_{21}}, \qquad \mathcal{B}_2 = \sum_{j=1}^{s} b_j^2 \frac{\partial}{\partial b_j} + (n+1+\alpha)\mathcal{B}_1 - c\frac{\partial}{\partial c_{12}}.$$

Note that $\mathcal{A}_2$ and $\mathcal{B}_2$ acting on $\tau_n$ depends on the index $n$.

THEOREM 12.1 (Laguerre distibution). *The joint statistics* (11.2), *namely $P_n(E) = \tau_n(E)/\tau_n(\mathbb{R}^+)$, is a ratio of two functions, each satisfying the non-linear third-order partial differential equation*[19] *($G_n := \log \tau_n(E)$):*

(12.4)
$$\left\{(\mathcal{B}_2\mathcal{A}_1 + c\frac{\partial}{\partial c_{12}})G_n \; , \; \mathcal{B}_1\mathcal{A}_1 G_n\right\}_{\mathcal{A}_1} - \left\{(\mathcal{A}_2\mathcal{B}_1 + c\frac{\partial}{\partial c_{21}})G_n \; , \; \mathcal{A}_1\mathcal{B}_1 G_n\right\}_{\mathcal{B}_1} = 0 \, .$$

*Remark.* Equation (12.4) is actually an inductive set of equations with regard to $n$, since it contains derivatives of the form $\partial \tau_n(E)/\partial c_{21}$ and $\partial \tau_n(E)/\partial c_{12}$. The point is that, according to (10.24), these derivatives can be expressed in terms of $(t,s)$-derivatives of the expression $\tau_{n-1}(E)$ in (12.5) below; to be precise,

$$\frac{\partial \tau_n(E)}{\partial c_{21}} = p_{n+1}(\tilde{\partial}_t)p_n(-\tilde{\partial}_s)\tau_1(E) \circ \tau_{n-1}(E)\Big|_{t=s=c_{ij}=0, c_{11}=c}$$
$$= \sum_{\substack{i+i'=n+1 \\ j+j'=n \\ i,i',j,j'\geq 0}} \mu_{ij}^E p_{i'}(-\tilde{\partial}_t)p_{j'}(\tilde{\partial}_s)\tau_{n-1}(E)\Big|_{t=s=c_{ij}=0, c_{11}=c} \, .$$

The $t,s$-partials of $\tau_{n-1}(E)$ can then be expressed again in terms the operators $\mathcal{A}_i, \mathcal{B}_i$ applied to $\tau_{n-1}$, etc... .

This result hinges on knowing, as before, the Virasoro constraints for the $(t,s)$ deformations of the matrix integral (12.1). Unlike the Gaussian case, which could be obtained by merely shifting the time, we invoke here a method, due to [5], of representing the matrix integral by vertex operators acting on a vacuum vector.

---

[19]in terms of the Wronskian $\{f,g\}_X = Xf.g - f.Xg$, with regard to a first-order differential operator $X$.

PROPOSITION 12.2. *The integral*

(12.5)
$$\tau_n(E) = \iint_{E^n} dxdy \Delta_n(x)\Delta_n(y) \prod_{k=1}^n \rho(x_k)\tilde{\rho}(y_k) e^{\sum_{i=1}^\infty (t_i x_k^i - s_i y_k^i) + \sum_{i,j\geq 1} c_{ij} x_k^i y_k^j},$$

*with the weights $\rho$ and $\tilde{\rho}$ satisfying*

(12.6) $\quad -\dfrac{\rho'}{\rho} = \dfrac{g}{f} = \dfrac{\sum_{i\geq 0}\beta_i z^i}{\sum_{i\geq 0}\alpha_i z^i} \quad and \quad -\dfrac{\tilde{\rho}'}{\tilde{\rho}} = \dfrac{\tilde{g}}{\tilde{f}} = \dfrac{\sum_{i\geq 0}\tilde{\beta}_i z^i}{\sum_{i\geq 0}\tilde{\alpha}_i z^i},$

*satisfies the following Virasoro equations for $k \geq -1$:*
(12.7)
$$\left(\sum_1^{2r} a_i^{k+1} f(a_i)\frac{\partial}{\partial a_i} - \sum_{i\geq 0}\left(\alpha_i(\mathbb{J}^{(2)}_{k+i} + \sum_{m,\ell\geq 1} mc_{m\ell}\frac{\partial}{\partial c_{m+k+i,\ell}}) - \beta_i \mathbb{J}^{(1)}_{k+i+1}\right)\right)\tau = 0$$

$$\left(\sum_1^{2r} b_i^{k+1} \tilde{f}(b_i)\frac{\partial}{\partial b_i} - \sum_{i\geq 0}\left(\tilde{\alpha}_i(\tilde{\mathbb{J}}^{(2)}_{k+i} + \sum_{m,\ell\geq 1} \ell c_{m\ell}\frac{\partial}{\partial c_{m,\ell+k+i}}) - \tilde{\beta}_i \tilde{\mathbb{J}}^{(1)}_{k+i+1}\right)\right)\tau = 0.$$

*Proof.* See [5].

*Proof of Theorem* 12.1. Since
$$\rho(z) = \tilde{\rho}(z) = z^\alpha e^{-z},$$
with
$$-\frac{\rho'}{\rho} = \frac{z-\alpha}{z} \quad and \quad f(z) = z,$$
we have $\beta_0 = -\alpha$, $\beta_1 = 1$, $\alpha_0 = 0$, $\alpha_1 = 1$ and all remaining $\alpha_i$ and $\beta_i = 0$. Therefore, for all $k \geq 0$,

(12.8) $\quad \left(\sum_1^{2r} a_i^{k+1}\frac{\partial}{\partial a_i} - \left(\mathbb{J}^{(2)}_k + \alpha\mathbb{J}^{(1)}_k - \mathbb{J}^{(1)}_{k+1} + \sum_{i,j\geq 1} ic_{ij}\frac{\partial}{\partial c_{k+i,j}}\right)\right)\tau = 0$

$$\left(\sum_1^{2r} b_i^{k+1}\frac{\partial}{\partial b_i} - \left(\tilde{\mathbb{J}}^{(2)}_k + \alpha\tilde{\mathbb{J}}^{(1)}_k - \tilde{\mathbb{J}}^{(1)}_{k+1} + \sum_{i,j\geq 1} jc_{ij}\frac{\partial}{\partial c_{i,j+k}}\right)\right)\tau = 0,$$

where $\mathbb{J}^{(2)}_k, \tilde{\mathbb{J}}^{(2)}_k, \mathbb{J}^{(1)}_k, \tilde{\mathbb{J}}^{(1)}_k$ are defined in (10.2) and in the formulas following Theorem 3.1. Guided by the same principle as (11.8), one redefines

$$\mathcal{V}_1 = \sum_1^\infty it_i\frac{\partial}{\partial t_i} - \frac{\partial}{\partial t_1} + \sum_{\substack{i,j\geq 1 \\ (i,j)\neq(1,1)}} ic_{ij}\frac{\partial}{\partial c_{ij}} + \frac{n(n+2\alpha+1)}{2}$$

$$\mathcal{V}_2 = \sum_1^\infty it_i \frac{\partial}{\partial t_{i+1}} - \frac{\partial}{\partial t_2} + \sum_{\substack{i,j \geq 1 \\ (i,j) \neq (1,1)}} ic_{ij} \frac{\partial}{\partial c_{i+1,j}} + (n+\alpha+1)(\mathcal{V}_1 + \frac{\partial}{\partial t_1})$$

and similarly for $\mathcal{W}_i$, using the map $s \leftrightarrow -t$ and $ic_{ij}\partial/\partial c_{k+i,j} \leftrightarrow jc_{ij}\partial/\partial c_{i,j+k}$. Here the involution acts as follows:

$$\mathcal{A}_i \leftrightarrow \mathcal{B}_i, \quad \mathcal{V}_i \leftrightarrow \mathcal{W}_i.$$

Then $\tau_n$ satisfies for $k = 1, 2$

(12.9) $$\mathcal{A}_k \tau_n = \mathcal{V}_k \tau_n \text{ and } \mathcal{B}_k \tau_n = \mathcal{W}_k \tau_n.$$

Evaluating $\mathcal{A}_k \tau_n$, $\mathcal{B}_k \tau_n$, $\mathcal{A}_1 \mathcal{B}_1 \tau_n$, $\mathcal{A}_2 \mathcal{B}_1 \tau_n$ along the locus $\mathcal{L}$, using the commutation relation $[\mathcal{A}_2, \mathcal{W}_1]|_\mathcal{L} = -c\frac{\partial}{\partial c_{21}}$ and (12.9), and setting $d_n = n(n + 2\alpha + 1)/2$, one checks

$$\left.\frac{\partial \tau_n}{\partial t_1}\right|_\mathcal{L} = -(\mathcal{A}_1 - d_n)\tau_n|_\mathcal{L} \quad \left.\frac{\partial \tau_n}{\partial s_1}\right|_\mathcal{L} = (\mathcal{B}_1 - d_n)\tau_n|_\mathcal{L}$$

$$\left.\frac{\partial^2 \tau_n}{\partial s_1 \partial t_1}\right|_\mathcal{L} = -(\mathcal{A}_1 - d_n)(\mathcal{B}_1 - d_n)\tau_n|_\mathcal{L}$$

$$\left.\frac{\partial^2 \tau_n}{\partial s_1 \partial t_2}\right|_\mathcal{L} = -(\mathcal{A}_2 - (n+\alpha+1)d_n)(\mathcal{B}_1 - d_n)\tau_n - c\frac{\partial}{\partial c_{21}}\tau_n\Big|_\mathcal{L}$$

and so,

$$\left.\frac{\partial \log \tau_n}{\partial s_1}\right|_\mathcal{L} = \mathcal{B}_1 \log \tau_n|_\mathcal{L} - \frac{n(n+2\alpha+1)}{2}$$

$$\left.\frac{\partial^2 \log \tau_n}{\partial s_1 \partial t_1}\right|_\mathcal{L} = -\mathcal{A}_1 \mathcal{B}_1 \log \tau_n|_\mathcal{L} \text{ and } \left.\frac{\partial^2 \log \tau_n}{\partial s_1 \partial t_2}\right|_\mathcal{L} = -\left(\mathcal{A}_2\mathcal{B}_1 + c\frac{\partial}{\partial c_{21}}\right)\log \tau_n\Big|_\mathcal{L}.$$

Setting these expressions into equations (11.14), we find an equation and its dual:

$$-\mathcal{A}_1 \log \frac{\tau_{n+1}}{\tau_{n-1}} + (2n + 2\alpha + 1) = \frac{(\mathcal{A}_2\mathcal{B}_1 + c\frac{\partial}{\partial c_{21}})\log \tau_n}{\mathcal{A}_1 \mathcal{B}_1 \log \tau_n}$$

$$-\mathcal{B}_1 \log \frac{\tau_{n+1}}{\tau_{n-1}} + (2n + 2\alpha + 1) = \frac{(\mathcal{B}_2\mathcal{A}_1 + c\frac{\partial}{\partial c_{12}})\log \tau_n}{\mathcal{B}_1 \mathcal{A}_1 \log \tau_n}.$$

Finally, upon subtracting $\mathcal{B}_1$ of the first from $\mathcal{A}_1$ of the second, one is led to:

$$\mathcal{B}_1 \frac{(\mathcal{A}_2\mathcal{B}_1 + c\frac{\partial}{\partial c_{21}})\log \tau_n}{\mathcal{A}_1 \mathcal{B}_1 \log \tau_n} - \mathcal{A}_1 \frac{(\mathcal{B}_2\mathcal{A}_1 + c\frac{\partial}{\partial c_{12}})\log \tau_n}{\mathcal{B}_1 \mathcal{A}_1 \log \tau_n} = 0,$$

ending the proof of Theorem 12.1. □

976 M. ADLER AND P. VAN MOERBEKE

BRANDEIS UNIVERSITY, WALTHAM, MA 02454, USA
*E-mail address*: adler@math.brandeis.edu

UNIVERSITÉ DE LOUVAIN, 1348 LOUVAIN-LA-NEUVE, BELGIUM, AND
BRANDEIS UNIVERSITY, WALTHAM, MA 02454, USA
*E-mail address*: vanmoerbeke@geom.ucl.ac.be and vanmoerbeke@math.brandeis.edu

## References

[1] M. ADLER AND P. VAN MOERBEKE, Matrix integrals, Toda symmetries, Virasoro constraints and orthogonal polynomials, Duke Math. J. **80** (1995), 863–911.

[2] ———, String-orthogonal polynomials, string equations and 2-Toda symmetries, Comm. Pure and Appl. Math. J. **50** (1997), 241–290.

[3] ———, Group factorization, moment matrices and 2-Toda lattices, Intern. Math. Research Notices **12** (1997), 555–572.

[4] ———, Vertex operator solutions to the discrete KP hierarchy, Comm. Math. Phys. **203** (1999), 185–210.

[5] ———, On beta-integrals, preprint (1999).

[6] M. ADLER, T. SHIOTA, and P. VAN MOERBEKE, From the $w_\infty$-algebra to its central extension: a $\tau$-function approach, Physics Lett. A **194** (1994) 33–43, and A Lax pair representation for the Vertex operator and the central extension, Comm. Math. Phys. **171** (1995), 547–588.

[7] ———, Random matrices, vertex operators and the Virasoro algebra, Phys. Lett. A **208** (1995), 67–78, and Random matrices, Virasoro algebras and "non-commutative" KP, Duke Math. J. **94** (1998), 379–431.

[8] D. BESSIS, CL. ITZYKSON, and J.-B. ZUBER, Quantum field theory techniques in graphical enumeration, Adv. in Appl. Math. **1** (1980), 109–157.

[9] J. J. DUISTERMAAT and G. HECKMAN, On the variation in the cohomology of the symplectic form of the reduced phase space, Invent. Math. **69** (1982), 259–268.

[10] F. DYSON, Fredholm determinants and inverse scattering problems, Comm. Math. Phys. **47** (1976), 171–183.

[11] B. EYNARD and M. L. MEHTA, Matrices coupled in a chain: I. Eigenvalue correlations, J. Phys. A: Math. Gen. **31** (1998), 4449-4456.

[12] HARISH CHANDRA, Differential operators on a semi-simple Lie algebra, Amer. J. of Math. **79** (1957), 87–120.

[13] A. R. ITS, A. G. IZERGIN, V. E. KOREPIN, and N. A. SLAVNOV, Differential equations for quantum correlation functions, Internat. J. Mod. Phys. B **4** (1990), 1003–1037.

[14] M. JIMBO, T. MIWA, Y. MÔRI, and M. SATO, Density matrix of an impenetrable Bose gas and the fifth Painlevé transcendent, Physica D **1** (1980), 80–158.

[15] G. MAHOUX, M. L. MEHTA, and J. M. NORMAND, Matrices coupled in a chain: II. Spacing functions, J. Phys. A: Math. Gen. **31** (1998), 4457–4464.

[16] M. L. MEHTA, *Random Matrices*, 2nd ed., Academic Press, Boston, 1991.

[17] M. L. MEHTA and P. SHUKLA, Two coupled matrices: Eigenvalue correlations and spacing functions, J. Phys. A: Math. Gen. **27** (1994) 7793–7803.

[18] K. UENO and K. TAKASAKI, Toda Lattice Hierarchy, Adv. Studies Pure Math. **4** (1984), 1–95.

[19] J. W. VAN DE LEUR, The $W_{1+\infty}(gl_s)$-symmetries of the $s$-component KP hierarchy, J. of Math. Phys. **37** (1996), 2315–2337.

[20] P. VAN MOERBEKE, The spectrum of random matrices and integrable systems, Group 21, *Physical Applications and Mathematical Aspects of Geometry, Groups, and Algebras*, Vol. II (H.-D. Doebner, W. Scherer, and C. Schulte, eds.), World Scientific Publ., Singapore, 1997.